\documentclass[11pt,draft]{llncs}
\usepackage{latexsym,amsfonts,amsmath,amssymb}
\usepackage{biograph}
\usepackage{mathptmx}
\usepackage{amssymb}
\usepackage{amsfonts}
\usepackage{amsmath}
\usepackage{latexsym}
\renewcommand{\leq}{\leqslant}
\renewcommand{\geq}{\geqslant}
\newcommand{\CC}{\mathcal{C}}
\renewcommand{\AA}{\mathcal{A}}

\newcommand{\XX}{\mathcal{X}}
\newcommand{\LL}{\mathcal{L}}
\newcommand{\cami}{\!\rightsquigarrow\!}
\usepackage[algo2e,boxed,vlined]{algorithm2e}
\usepackage{color}
\definecolor{lightgray}{gray}{0.75}
\usepackage{listings}
\renewcommand{\qed}{\hbox{}\nobreak\hfill\vrule width 1.4mm height 1.4mm depth 0mm
    \par \goodbreak \smallskip}

\begin{document}

\title{On the Ancestral Compatibility of Two Phylogenetic Trees with
  Nested Taxa}

    \author{Merc\`e Llabr\'es\inst{1}, Jairo Rocha\inst{1}, Francesc Rossell\'o\inst{1} \and Gabriel Valiente\inst{2}}

    \institute{Department of Mathematics and Computer Science,
    Research Institute of Health Science (IUNICS), University of the
    Balearic Islands, E-07122 Palma de Mallorca,
    \{\texttt{merce.llabres,jairo.rocha,cesc.rossello}\}\texttt{@uib.es} \and Department of Software,
    Technical University of Catalonia, E-08034 Barcelona,
    \texttt{valiente@lsi.upc.es}}

    \maketitle

    \begin{abstract}
      Compatibility of phylogenetic trees is the most important
      concept underlying widely-used methods for assessing the
      agreement of different phylogenetic trees with overlapping taxa
      and combining them into common supertrees to reveal the tree of
      life.  The notion of ancestral compatibility of phylogenetic
      trees with nested taxa was introduced by Semple et al in 2004.
      In this paper we analyze in detail the meaning of this
      compatibility from the points of view of the local structure of
      the trees, of the existence of embeddings into a common
      supertree, and of the joint properties of their cluster
      representations.  Our analysis leads to a very simple
      polynomial-time algorithm for testing this compatibility, which
      we have implemented and is freely available for download from
      the BioPerl collection of Perl modules for computational
      biology.
    \end{abstract}


    \section{Introduction}

    A rooted phylogenetic tree can be seen as a static description of the
    evolutive history of a family of contemporary species: these species
    are located at the leaves of the tree, and their common ancestors are
    organized as the inner nodes of the tree. These interior nodes
    represent taxa at a higher level of aggregation or nesting than that
    of their descendents, ranging for instance from families over genera
    to species. Phylogenetic trees with nested taxa have thus all leaves
    as well as some interior nodes labeled, and they need not be
    fully-resolved trees and may have unresolved polytomies, that is, they
    need not be binary trees.

    Often one has to deal with two or more phylogenetic trees with
    overlapping taxa, probably obtained through different techniques by
    the same or different researchers. The problem of combining these
    trees into a single supertree containing the evolutive information of
    all the given trees has recently received much attention, and it has
    been identified as a promising approach to the reconstruction of the
    tree of life~\cite{bininda:2004}. This information corresponds to
    evolutive precedence, and hence it is kept when every arc in each of
    the trees becomes a path in the supertree.

    It is well known that it is not always possible to combine
    phylogenetic trees into a single supertree: there are
    \emph{incompatible} phylogenetic trees that do not admit their
    simultaneous inclusion into a common supertree.  Compatibility for
    leaf-labeled phylogenetic trees was first studied
    in~\cite{warnow:1994}.  Incompatible phylogenetic trees can still be
    partially combined into a maximum agreement
    subtree~\cite{steel.warnow:1993}.  Compatible phylogenetic trees, on
    the other hand, can be combined into a common supertree, two of the
    most widely used methods being matrix representation with
    parsimony~\cite{baum:1992,ragan:1992} and
    mincut~\cite{page:2002,semple.ea:2000} and it is clear that, because
    of Occam's razor, one is interested in obtaining not only a common
    supertree of the given phylogenetic trees, but the smallest possible
    one.  The relationship between the largest common subtree and the
    smallest common supertree of two leaf-labeled phylogenetic trees was
    established in~\cite{UPC-LSI-04-60-R} by means of simple
    constructions, which allow one to obtain the largest common subtree
    from the smallest common supertree, and vice versa.

    The study of the compatibility of phylogenetic trees with nested taxa,
    also known as \emph{semi-labeled trees}, was asked for in
    \cite{page:2004}. Polynomial-time algorithms were proposed
    in~\cite{daniel.semple:2004,semple.ea:2004} for testing a weak form of
    compatibility, called \emph{ancestral compatibility}, and a stronger
    form called \emph{perfect compatibility}.  Roughly, two or more
    semi-labeled trees are ancestrally compatible if they can be refined
    into a common supertree, and they are perfectly compatible if there
    exists a common supertree whose topological restriction to the taxa in
    each tree is isomorphic to that tree.

    In this paper, we are concerned with the notion of ancestral
    compatibility of semi-labeled trees. In particular, we establish the
    equivalence between this notion and the absence of certain
    `incompatible' pairs and triples of labels in the trees under
    comparison.  We also prove the equivalence between ancestral
    compatibility and a certain property of the cluster representations of
    the trees.  These equivalences lead to a new polynomial-time algorithm
    for testing ancestral compatibility of semi-labeled trees, which we
    have implemented and is freely available for download from the BioPerl
    collection of Perl modules for computational
    biology~\cite{stajich.ea:2002}.

    %

    The rest of the paper is organized as follows.  Basic notions and
    notation are recalled in Section~\ref{sect:prelim}.  A notion of local
    compatibility as the absence of incompatible pairs and triples of
    labels is introduced in Section~\ref{Atrees}, together with some basic
    results about a relaxed notion of semi-labeled trees.  Weak
    topological embeddings, and the notion of ancestral compatibility that
    derives from them, are studied in Section~\ref{sect:embed}.  In
    Section~\ref{sect:main}, the equivalence between local compatibility
    in the sense of Section~\ref{Atrees} and ancestral compatibility in
    the sense of Section~\ref{sect:embed} is established, as well as a
    characterization in terms of cluster representations. The BioPerl
    implementation of the algorithm for testing compatibility of two
    semi-labeled trees is described in Section~\ref{sect:implement}.
    Finally, some conclusions and further work are outlined in
    Section~\ref{sect:concl}.

    \section{Preliminaries}
    \label{sect:prelim}

    Throughout this paper, by a \emph{tree} we mean a \emph{rooted tree},
    that is, a directed finite graph $T=(V,E)$ with $V$ either empty or
    containing a distinguished node $r\in V$, called the \emph{root}, such
    that for every other node $v\in V$ there exists one, and only one,
    path from the root $r$ to $v$.  Recall that every node in a tree has
    in-degree 1, except the root, which has in-degree 0.

    Henceforth, and unless otherwise stated, given a tree $T$ we shall
    denote its set of nodes by $V(T)$ and its set of arcs by $E(T)$.  
    The \emph{children} of a node $v$ in a tree $T$ are those nodes $w$
    such that $(v,w)\in E(T)$.  The nodes without children are the
    \emph{leaves} of the tree, and we shall call \emph{elementary} the
    nodes with only one child.  

    Given a path $(v_{0},v_{1},\ldots,v_{k})$ in a tree $T$, its
    \emph{origin} is $v_{0}$, its \emph{end} is $v_{k}$, and its
    \emph{intermediate nodes} are $v_{1},\ldots,v_{k-1}$.  Such a path is
    \emph{non-trivial} when $k\geq 1$.  We shall represent a path
    \emph{from $v$ to $w$}, that is, a path with origin $v$ and end $w$,
    by $v\rightsquigarrow w$.  When there exists a path $v\rightsquigarrow
    w$, we say that $w$ is a \emph{descendant} of $v$ and also that $v$ is
    an \emph{ancestor} of $w$. Every node is both an ancestor and a
    descendant of itself, through a trivial path.

    Two non-trivial paths $(a,v_{1},\ldots,v_{k})$ and
    $(a,w_{1},\ldots,w_{\ell})$ in a tree $T$ are said to \emph{diverge}
    when the only node they have in common is their origin $a$.  Notice
    that, by the uniqueness of paths in trees, it is equivalent to the
    condition $v_{1}\neq w_{1}$.  For every two nodes $v,w$ of a tree that
    are not connected by a path, there exists one, and only one, common
    ancestor $a$ of $v$ and $w$ such that there exist divergent paths from
    $a$ to $v$ and to $w$.  We shall call it the \emph{most recent common
      ancestor} of $v$ and $w$.  When there is a path $v\cami w$, we say
    that $v$ is the \emph{most recent common ancestor} of $v$ and $w$.

    \section{$\AA$-trees}
    \label{Atrees}

    Let $\AA$ be throughout this paper a fixed set of labels.  In
    practice, we shall use the first capital letters, $A,B,C\ldots$, as
    labels. 

    \begin{definition}
    A  \emph{semi-labeled tree over $\AA$} is a tree with some of its
    nodes, including all its leaves and all its elementary nodes,
    injectively labeled in the set $\AA$.
    \end{definition}

    To simplify several proofs, we shall usually allow the existence of
    unlabeled elementary nodes. This motivates the following definition.

    \begin{definition}
    An \emph{$\AA$-tree} is a tree with some of its nodes, including all
    its leaves, injectively labeled in the set $\AA$.
    \end{definition}

    We shall always use the same name to denote an $\AA$-tree and the
    (unlabeled) tree that \emph{supports} it.  Furthermore, for every
    $\AA$-tree $T$, we shall use henceforth the following notations:

    \begin{itemize}

    \item $\LL(T)$ and $\AA(T)$ will denote, respectively, the set of the labels
    of its leaves and the set of the labels of all its nodes.


    \item For every $v\in V(T)$, we shall denote by $\AA_{T}(v)$ the set
      of the labels of all its descendants, including itself, and we shall
      call it, following \cite{semple.steel:2003}, the \emph{cluster} of
      $v$ in $T$; if $T$ is irrelevant or clearly determined by the
      context, we shall usually write $\AA(v)$ instead of $\AA_{T}(v)$.
      Notice that if there exists a path $w\cami v$, then $\AA(v)\subseteq
      \AA(w)$.

    \item
    We shall set
    \[
    \CC_{\AA}(T)=\{\AA_{T}(v)\mid v\in V(T)\}.
    \]
    Notice that $\emptyset\notin\CC_{\AA}(T)$ unless $T$ is empty. If $T$
    is a semi-labeled tree over $\AA$, then $\CC_{\AA}(T)$ coincides with
    the cluster representation~\cite{semple.steel:2003} of $T$, up to the
    trivial cluster for the root of $T$. Consequently, even for $\AA$-trees, 
    we shall call $\CC_{\AA}(T)$ the \emph{cluster representation} of $T$.

    \item For every $X\subseteq \AA(T)$, we shall denote by $v_{T,X}$ the
      most recent common ancestor of the nodes of $T$ with labels in $X$;
      when $T$ is irrelevant or clearly determined by the context, we
      shall usually write $v_{X}$ instead of $v_{T,X}$.  Moreover, when
      $X$ is given by the list of its members between brackets, we shall
      usually omit these brackets in the subscript.  So, in particular,
      for every $A\in \AA(T)$, we shall denote the node of $T$ labeled $A$
      by $v_{T,A}$ or simply $v_{A}$.

    Notice that $\AA(v_{T,X})=X$ if and only if $X\in \CC_{\AA}(T)$.

    \end{itemize}

    We shall often use the following easy results, usually without any
    further mention.

    \begin{lemma}\label{lem-int1}
    Let $T$ be an $\AA$-tree, and let $x,y\in V(T)$.  If
    $\AA(x)\cap\AA(y)\neq\emptyset$, then $x$ is a descendant of $y$ or
    $y$ is a descendant of $x$.
    \end{lemma}

    \begin{proof}
    Let $A\in \AA(x)\cap \AA(y)$, so that there exist paths $x\cami
    v_{A}$ and $y\cami v_{A}$, and let $r$ be the root of $T$.  Then,
    both $x$ and $y$ appear in the path $r\cami v_{A}$.  This entails
    that either $x$ appears in the path $y\cami v_{A}$ or $y$ appears in
    the path $x\cami v_{A}$, meaning that there is either a path from
    $y$ to $x$ or from $x$ to $y$.
    \qed
    \end{proof}

    \begin{corollary}\label{lem-int2}
    Let $T$ be an $\AA$-tree, and let $x,y\in V(T)$.  If
    $\AA(x)\subsetneq\AA(y)$, then there is a non-trivial path $y\cami x$.
    \end{corollary}

    \begin{proof}
    By the previous lemma, if $\AA(x)\subsetneq \AA(y)$, then either $x$
    is a descendant of $y$ or $y$ is a descendant of $x$.  But, being the
    inclusion strict, $y$ cannot be a descendant of $x$.  
    \qed
    \end{proof}

    \begin{corollary}\label{lem-int3}
    Let $T$ be an $\AA$-tree, and let $x,y\in V(T)$ be two different
    nodes.  If $\AA(x)=\AA(y)$, then there is a path $x\cami y$ or a path
    $y\cami x$, such that its origin and all its intermediate nodes are
    unlabeled and elementary.
    \end{corollary}

    \begin{proof}
    By Lemma~\ref{lem-int1}, if $\AA(x)=\AA(y)$, there is either a path
    $x\cami y$ or a path $y\cami x$.  If the origin or some intermediate
    node in this path is labeled or if any one of these nodes has more
    children that those appearing in this path, then the set of labels
    will decrease from this node to its child in the path, and \textsl{a
    fortiori} from the origin to the end of the path.
    \qed
    \end{proof}

    In particular, in a semi-labeled tree over $\AA$, which does not
    contain any unlabeled elementary node, $\AA(x)=\AA(y)$ if and only if
    $x=y$, and $\AA(x)\subsetneq \AA(y)$ if and only if there exists a
    non-trivial path $y\cami x$.  This entails that the cluster
    representation $\CC_{\AA}(T)$ of a semi-labeled tree $T$ over $\AA$
    determines $T$ up to
    isomorphism~\cite[Theorem~3.5.2]{semple.steel:2003}.

    \begin{definition}
    The \emph{restriction} $T|\XX$ of an $\AA$-tree $T$ to a set
    $\XX\subseteq \AA$ of labels is the subtree of $T$ supported on the
    set of nodes 
    $$
    \begin{array}{rl}
    V(T|\XX) & = \{v\in V(T)\mid \mbox{there exists a path $v\cami
    v_{A}$ for some $A\in\XX$}\}\\
    & = \{v\in V(T)\mid \AA(v)\cap \XX\neq \emptyset\},
    \end{array}
    $$ 
    and where a node is labeled when it is labeled in $T$ and
    this label belongs to $\XX$, in which case
    its label in $T|\XX$ is the same as in $T$.
    \end{definition}

    If $\XX\cap\AA(T)=\emptyset$, then $T|\XX$ is the empty $\AA$-tree,
    while if $\XX\cap\AA(T)\neq\emptyset$, then $T|\XX$ has the same root
    as $T$ and leaves the nodes of $T$ with labels in $\XX$ that do not
    have any descendant with label in $\XX$. 

    Now we introduce the notion of \emph{locally compatible $\AA$-trees} as the
    absence of \emph{incompatible pairs and triples} of labels.

    \begin{definition}
    Two $\AA$-trees $T_{1}$ and $T_{2}$ are \emph{locally compatible} when
    they satisfy the following two conditions:
    \begin{description}
    \item[(C1)] For every two labels $A,B\in \AA(T_{1})\cap \AA(T_{2})$,
    there is a path $v_{A}\cami v_{B}$ in $T_{1}$ if and only
    if there is a path $v_{A}\cami v_{B}$ in $T_{2}$.

    \item[(C2)] For every three labels $A,B,C\in \AA(T_{1})\cap
    \AA(T_{2})$, if there exists a non-trivial path $v_{B,C}\cami v_{A,B}$
    in $T_{1}$, then there does not exist any non-trivial path $v_{A,B}\cami
    v_{B,C}$ in $T_{2}$.
    \end{description}
    Any pair of labels $A,B$ violating condition (C1) and any triple of labels 
    $A,B,C$ violating condition (C2) in a pair of trees $T_{1}$ and $T_{2}$
    are said to be \emph{incompatible}.

    Two $\AA$-trees $T_{1}$ and $T_{2}$ are \emph{locally incompatible}
    when they are not locally compatible, that is, when they contain an
    incompatible pair or triple of labels.
    \end{definition}

    So, if $T_{1}$ and $T_{2}$ represent phylogenetic trees with nested
    taxa, an incompatible pair of labels in $T_{1}$ and $T_{2}$
    corresponds to a pair of taxa whose evolutive precedence is different
    in both trees, while an incompatible triple of labels in $T_{1}$ and
    $T_{2}$ corresponds to three  taxa whose evolutive divergence is
    different in both trees.

\begin{figure}[htb]
\begin{center}
        \begin{picture}(140,40)(0,0)
        \put(0,0){\line(1,1){40}} 
        \put(80,0){\line(-1,1){40}} 
        \put(0,0){\circle*{3}} 
        \put(20,20){\circle*{3}} 
        \put(40,40){\circle*{3}} 
        \put(80,0){\circle*{3}} 
        \put(-5,0){\makebox(0,0)[r]{$A$}}
        \put(15,20){\makebox(0,0)[r]{$B$}}
        \put(85,0){\makebox(0,0)[l]{$C$}}
        \put(65,25){\makebox(0,0)[l]{$T_{1}$}}
        \put(120,0){\line(0,1){40}} 
        \put(120,0){\circle*{3}} 
        \put(120,20){\circle*{3}} 
        \put(120,40){\circle*{3}} 
        \put(125,0){\makebox(0,0)[l]{$A$}}
        \put(125,20){\makebox(0,0)[l]{$B$}}
        \put(125,40){\makebox(0,0)[l]{$C$}}
        \put(140,25){\makebox(0,0)[l]{$T_{1}$}}
        \end{picture}
        \caption{\label{fig:Inc-resta}%
        These two $\AA$-trees are only locally compatible with themselves}
\end{center}
    \end{figure}
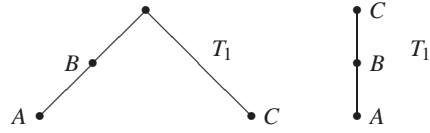

    \begin{example}\label{rem:ABCinc}
    Let $T_{1},T_{2}$ be two locally compatible $\AA$-trees, and let $A,B,C\in
    \AA(T_{1})\allowbreak\cap \AA(T_{2})$.  If $T_{1}$ contains a structure above
    $v_{A},v_{B},v_{C}$ as the one shown in the left-hand side of
    Fig.~\ref{fig:Inc3},\footnote{In this figure, as well as in
    Figs.~\ref{fig:Inc3bis} to~\ref{fig:Inc-resta}, edges may represent
    actually non-trivial paths.} then $T_{2}$ contains either the same
    structure above $v_{A},v_{B},v_{C}$ as $T_{1}$ or the one shown in the
    right-hand side of the same figure.

    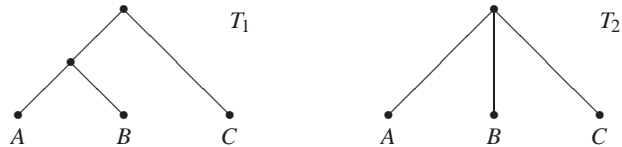
\begin{figure}[htb]
\begin{center}
        \begin{picture}(220,60)(0,-10)
        \put(0,0){\line(1,1){40}} 
        \put(40,0){\line(-1,1){20}} 
        \put(40,40){\line(1,-1){40}} 
        \put(0,0){\circle*{3}} 
        \put(40,0){\circle*{3}} 
        \put(80,0){\circle*{3}} 
        \put(20,20){\circle*{3}} 
        \put(40,40){\circle*{3}} 
        \put(0,-5){\makebox(0,0)[t]{$A$}}
        \put(40,-5){\makebox(0,0)[t]{$B$}}
        \put(80,-5){\makebox(0,0)[t]{$C$}}
        \put(80,35){\makebox(0,0)[l]{$T_{1}$}}
        \put(140,0){\line(1,1){40}} 
        \put(180,0){\line(0,1){40}} 
        \put(220,0){\line(-1,1){40}} 
        \put(140,0){\circle*{3}} 
        \put(180,0){\circle*{3}} 
        \put(220,0){\circle*{3}} 
        \put(180,40){\circle*{3}} 
        \put(140,-5){\makebox(0,0)[t]{$A$}}
        \put(180,-5){\makebox(0,0)[t]{$B$}}
        \put(220,-5){\makebox(0,0)[t]{$C$}}
        \put(220,35){\makebox(0,0)[l]{$T_{2}$}}
        \end{picture}
        \caption{\label{fig:Inc3}%
        $T_{1}$ and $T_{2}$ are locally compatible}
\end{center}
    \end{figure}

    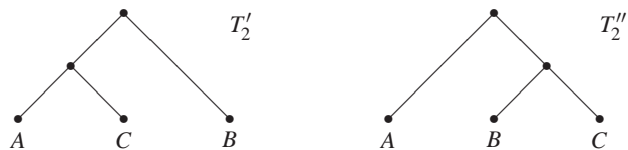
\begin{figure}[htb]
   \begin{center}
       \begin{picture}(220,60)(0,-10)
       \put(0,0){\line(1,1){40}} 
       \put(40,0){\line(-1,1){20}} 
       \put(40,40){\line(1,-1){40}} 
       \put(0,0){\circle*{3}} 
       \put(40,0){\circle*{3}} 
       \put(80,0){\circle*{3}} 
       \put(20,20){\circle*{3}} 
       \put(40,40){\circle*{3}} 
       \put(0,-5){\makebox(0,0)[t]{$A$}}
       \put(40,-5){\makebox(0,0)[t]{$C$}}
       \put(80,-5){\makebox(0,0)[t]{$B$}}
       \put(80,35){\makebox(0,0)[l]{$T_{2}'$}}
       \put(140,0){\line(1,1){40}} 
       \put(180,0){\line(1,1){20}} 
       \put(180,40){\line(1,-1){40}} 
       \put(140,0){\circle*{3}} 
       \put(180,0){\circle*{3}} 
       \put(220,0){\circle*{3}} 
       \put(200,20){\circle*{3}} 
       \put(180,40){\circle*{3}} 
       \put(140,-5){\makebox(0,0)[t]{$A$}}
       \put(180,-5){\makebox(0,0)[t]{$B$}}
       \put(220,-5){\makebox(0,0)[t]{$C$}}
       \put(220,35){\makebox(0,0)[l]{$T_{2}''$}}
       \end{picture}
       \caption{\label{fig:Inc3bis}%
       $T_{2}'$ and $T_{2}''$ are locally incompatible with $T_{1}$ in Fig.~\ref{fig:Inc3}}
   \end{center}
     \end{figure}

    Indeed, since no two among $v_{A},v_{B},v_{C}$ are connected in
    $T_{1}$ by a path, condition (C1) implies that no two among the nodes
    in $T_{2}$ labeled $A,B,C$ are connected by a path, either.  Beside
    the structures shown in Fig.~\ref{fig:Inc3}, only the structures
    $T_{2}'$ and $T_{2}''$ shown in Fig.~\ref{fig:Inc3bis} satisfy this
    property.  Now, $T_{1}$ contains a non-trivial path $v_{A,C}\cami
    v_{A,B}$, while $T_{2}'$ contains a non-trivial path $v_{A,B}\cami
    v_{A,C}$; and $T_{1}$ contains a non-trivial path $v_{B,C}\cami
    v_{A,B}$, while $T_{2}''$ contains a non-trivial path $v_{A,B}\cami
    v_{B,C}$. So, in both cases we find incompatible triples of labels.
    On the other hand, in the $\AA$-tree $T_{2}$ shown in
    Fig.~\ref{fig:Inc3}, $v_{A,B}= v_{A,C}= v_{B,C}$, and therefore this
    $\AA$-tree clearly satisfies condition (C2) with $T_{1}$ as far as the
    labels $A,B,C$ go.
    \end{example}


    \begin{example}\label{rem:A<-B->C}
    Let $T_{1},T_{2}$ be two locally compatible $\AA$-trees, and let
    $A,B,C\in \AA(T_{1})\allowbreak\cap
    \AA(T_{2})$.  If $T_{1}$ contains a structure above
    $v_{A},v_{B},v_{C}$ as the one shown in the
    left-hand side of Fig.~\ref{fig:Inc4}, then $T_{2}$ contains either the same structure above
    $v_{A},v_{B},v_{C}$ as $T_{1}$ or the one shown in
    the right-hand side of the same figure.

    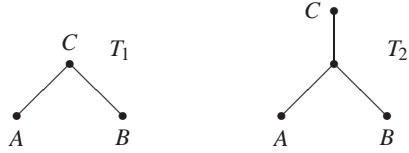
\begin{figure}[htb]
\begin{center}
        \begin{picture}(140,60)(0,-10)
        \put(100,0){\line(1,1){20}} 
        \put(140,0){\line(-1,1){20}} 
        \put(120,20){\line(0,1){20}} 
        \put(100,0){\circle*{3}} 
        \put(140,0){\circle*{3}} 
        \put(120,20){\circle*{3}} 
        \put(120,40){\circle*{3}} 
        \put(100,-5){\makebox(0,0)[t]{$A$}}
        \put(140,-5){\makebox(0,0)[t]{$B$}}
        \put(115,40){\makebox(0,0)[r]{$C$}}
        \put(140,25){\makebox(0,0)[l]{$T_{2}$}}
        \put(0,0){\line(1,1){20}} 
        \put(40,0){\line(-1,1){20}} 
        \put(0,0){\circle*{3}} 
        \put(40,0){\circle*{3}} 
        \put(20,20){\circle*{3}} 
        \put(0,-5){\makebox(0,0)[t]{$A$}}
        \put(40,-5){\makebox(0,0)[t]{$B$}}
        \put(20,25){\makebox(0,0)[b]{$C$}}
        \put(35,25){\makebox(0,0)[l]{$T_{1}$}}
        \end{picture}
        \caption{\label{fig:Inc4}%
        $T_{1}$ and $T_{2}$ are locally compatible}
\end{center}
    \end{figure}

    Indeed, in order to satisfy condition (C1), the existence in $T_{1}$
    of paths $v_{C}\cami v_{A}$, $v_{C}\cami v_{B}$ and the fact that
    $v_{A}$ and $v_{B}$ are not connected by a path in this $\AA$-tree,
    entail that $T_{2}$ also contains paths $v_{C}\cami v_{A}$,
    $v_{C}\cami v_{B}$ and that $v_{A}$ and $v_{B}$ are not connected by a
    path either.  Therefore, $T_{2}$ must either contain the same
    structure above $v_{A},v_{B},v_{C}$ as $T_{1}$, or 
    non-trivial paths $v_{C}\cami v_{A,B}$, $v_{A,B}\cami v_{A}$,
    $v_{A,B}\cami v_{B}$.  And since, in $T_{1}$,
    $v_{A,B}= v_{A,C}= v_{B,C}$, it is clear that in the last case the
    labels $A,B,C$ do not form an incompatible triple in $T_{1}$ and
    $T_{2}$.
    \end{example}

    \begin{example}\label{rem:altres}
    Let $T_{1},T_{2}$ be two locally compatible $\AA$-trees, and let
    $A,B,C\in \AA(T_{1})\allowbreak\cap \AA(T_{2})$.  If $T_{1}$ contains
    above $v_{A},v_{B},v_{C}$ one of the structures shown in
    Fig.~\ref{fig:Inc-resta}, then $T_{2}$ must contain the same structure
    above $v_{A},v_{B},v_{C}$.

    Indeed, it is a simple consequence of the application of condition
    (C1).  In the left-hand side structure, $T_{1}$ contains a path
    $v_{B}\cami v_{A}$, and $v_{B}$ and $v_{C}$
    are not connected by a path in it, and therefore the same must happen in
    $T_{2}$ and this leads to the same structure.
    And in the right-hand side structure, $T_{1}$ contains paths
    $v_{C}\cami v_{B}\cami v_{A}$, 
    and then the same must happen in
    $T_{2}$, entailing again the same structure in this tree.
    \end{example}

    The following construction will be used henceforth several times.

    \begin{definition}\label{def:hat}
    For every pair of $\AA$-trees $T_{1}$ and $T_{2}$, let
    $$
    \bar{T}_{1}=T_{1}|\AA(T_{1})\cap
    \AA(T_{2}),\quad\mbox{ and }\quad
    \bar{T}_{2}=T_{2}|\AA(T_{1})\cap \AA(T_{2}).
    $$
    \end{definition}

    Notice that, by construction, every leaf of each $\bar{T}_{i}$ is
    labeled, and therefore $\bar{T}_{1}$ and $\bar{T}_{2}$ are
    $\AA$-trees.  Notice also that if $\AA(T_{1})=\AA(T_{2})$, then
    $\bar{T}_{1}=T_{1}$ and $\bar{T}_{2}=T_{2}$.  In general,
    $$
    \AA(\bar{T}_{1})=\AA(\bar{T}_{2})=\AA(T_{1})\cap \AA(T_{2}).
    $$

    Since local compatibility of two $\AA$-trees refers to labels appearing
    in both $\AA$-trees, we clearly have the following result.

    \begin{lemma}\label{comp-hat}
    Two $\AA$-trees $T_{1}$ and $T_{2}$ are locally compatible if and only if
    $\bar{T}_{1}$ and $\bar{T}_{2}$ are so.  \qed
    \end{lemma}

    \section{Weak topological embeddings}
    \label{sect:embed}

    Compatibility of phylogenetic trees is usually stated in terms of the
    existence of simultaneous embeddings of some kind into a common supertree.  
    In this section we introduce the embeddings that will correspond to
    local compatibility.

    First, recall from \cite{semple.ea:2004} the definition of ancestral
    displaying, which we already present translated into our notations.

    \begin{definition}\label{def:disp}
    An $\AA$-tree $T$ \emph{ancestrally displays} an $\AA$-tree $S$ if the following properties
    hold:
    \begin{itemize}
    \item $\AA(S)\subseteq \AA(T)$.

    \item For every $A,B\in \AA(S)$, there is a path $v_{A}\cami
    v_{B}$ in $S$ if and only if there is a path $v_{A}\cami v_{B}$
    in $T$.

    \item $S$ is \emph{refined} by $T|\AA(S)$, that is,
    $\CC_{\AA}(S)\subseteq \CC_{\AA}(T|\AA(S))$.
    \end{itemize}
    \end{definition}

    We introduce now the following, more algebraic in flavour,
    definition of embedding that will turn out to be equivalent
    to ancestral displaying, up to the removal of elementary unlabeled
    nodes: cf.\ Proposition~\ref{prop:disp} below.

    \begin{definition}
      A \emph{weak topological embedding} of trees $f:S\to T$ is a mapping
      $f:V(S)\to V(T)$ satisfying the following conditions:
    \begin{itemize}
    \item It is \emph{injective}.
    \item It \emph{preserves labels}: for every $A\in \AA(S)$,
      $f(v_{A})=v_{A}$.
    \item It \emph{preserves and reflects paths}: for every $a,b\in V(S)$,
      there is a path from $a$ to $b$ in $S$ if and only if there is a
      path from $f(a)$ to $f(b)$ in $T$.
    \end{itemize}
    \end{definition}

    When a weak topological embedding of $\AA$-trees $f:S\to T$ exists, we
    say that $S$ is a \emph{weak $\AA$-subtree} of $T$ and that $T$ is a
    \emph{weak $\AA$-supertree} of $S$.

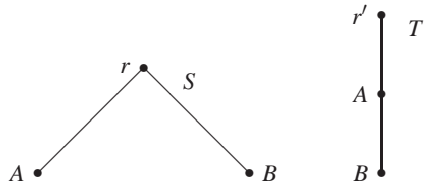
\begin{figure}[htb]
\begin{center}
        \begin{picture}(150,65)(-5,-10)
        \put(0,0){\line(1,1){40}} 
        \put(80,0){\line(-1,1){40}} 
        \put(0,0){\circle*{3}} 
        \put(80,0){\circle*{3}} 
        \put(40,40){\circle*{3}} 
        \put(-5,0){\makebox(0,0)[r]{$A$}}
        \put(85,0){\makebox(0,0)[l]{$B$}}
        \put(35,40){\makebox(0,0)[r]{$r$}}
        \put(55,35){\makebox(0,0)[l]{$S$}}
        \put(130,0){\line(0,1){60}} 
        \put(130,0){\circle*{3}} 
        \put(130,30){\circle*{3}} 
        \put(130,60){\circle*{3}} 
        \put(125,0){\makebox(0,0)[r]{$B$}}
        \put(125,30){\makebox(0,0)[r]{$A$}}
        \put(125,60){\makebox(0,0)[r]{$r'$}}
        \put(140,55){\makebox(0,0)[l]{$T$}}
        \end{picture}
        \caption{\label{fig:Inc5}%
          The $\AA$-trees in Example~\ref{ex:4-2}}
\end{center}
    \end{figure}

    \begin{example}
    \label{ex:4-2}
    Let $S$ and $T$ be the $\AA$-trees described in Fig.~\ref{fig:Inc5},
    and let $f:V(S)\to V(T)$ be the mapping defined by $f(r)=r'$,
    $f(v_{S,A})=v_{T,A}$ and $f(v_{S,B})=v_{T,B}$. This mapping is
    injective, preserves labels and preserves paths, but it does not
    reflect paths: there is a path $v_{T,A}\cami v_{T,B}$ is $T$, but no
    path from $v_{S,A}$ to $v_{S,B}$ in $S$. Therefore, it does not define
    a weak topological embedding $f:S\to T$.
    \end{example}

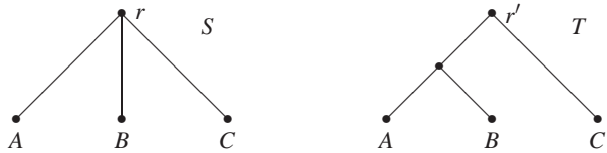
\begin{figure}[htb]
\begin{center}
        \begin{picture}(220,60)(0,-10)
        \put(140,0){\line(1,1){40}} 
        \put(180,0){\line(-1,1){20}} 
        \put(180,40){\line(1,-1){40}} 
        \put(140,0){\circle*{3}} 
        \put(180,0){\circle*{3}} 
        \put(220,0){\circle*{3}} 
        \put(160,20){\circle*{3}} 
        \put(180,40){\circle*{3}} 
        \put(185,40){\makebox(0,0)[l]{$r'$}}
        \put(140,-5){\makebox(0,0)[t]{$A$}}
        \put(180,-5){\makebox(0,0)[t]{$B$}}
        \put(220,-5){\makebox(0,0)[t]{$C$}}
        \put(210,35){\makebox(0,0)[l]{$T$}}
        \put(0,0){\line(1,1){40}} 
        \put(40,0){\line(0,1){40}} 
        \put(80,0){\line(-1,1){40}} 
        \put(0,0){\circle*{3}} 
        \put(40,0){\circle*{3}} 
        \put(80,0){\circle*{3}} 
        \put(40,40){\circle*{3}} 
        \put(45,40){\makebox(0,0)[l]{$r$}}
        \put(0,-5){\makebox(0,0)[t]{$A$}}
        \put(40,-5){\makebox(0,0)[t]{$B$}}
        \put(80,-5){\makebox(0,0)[t]{$C$}}
        \put(70,35){\makebox(0,0)[l]{$S$}}
        \end{picture}
        \caption{\label{fig:Inc3-repe}%
          The $\AA$-trees in Example~\ref{ex:4-1}}
\end{center}
    \end{figure}

    \begin{example}
    \label{ex:4-1}
    Let $S$ and $T$ the $\AA$-trees described in Fig.~\ref{fig:Inc3-repe}.
    Let $f:V(S)\to V(T)$ be the mapping that sends the root $r$ of $S$ to
    the root $r'$ of $T$, and every leaf of $S$ to the leaf of $T$ with
    the same label.  This mapping is injective, preserves labels, and
    preserves and reflects paths.  Therefore, it is a weak topological
    embedding $f:S\to T$.
    \end{example}

    \begin{example}
    \label{ex:restr-subt}
    For every $\AA$-tree $T$ and for every $\XX\subseteq \AA(T)$, the
    inclusion of the restriction $T|\XX$ into $T$ is a weak topological
    embedding.
    \end{example}

    \begin{remark}
    \label{pres-paths}
    It is straightforward to prove that a mapping $f:V(S)\to V(T)$
    preserves paths if and only if it \emph{transforms arcs into paths},
    that is, for every $a,b\in V(S)$, if $(a,b)\in E(S)$, then there
    exists a path $f(a)\cami f(b)$ in $T$.  We shall sometimes use this
    alternative formulation without any further mention.
    \end{remark}

    The following lemmas will be used several times in the sequel.

    \begin{lemma}
    \label{Av=Af(v)}
    Let $f:S\to T$ be a weak topological embedding.  Then, for every $v\in
    V(S)$, $\AA(v)=\AA(f(v))\cap \AA(S)$.
    \end{lemma}

    \begin{proof}
      The inclusion $\AA(v)\subseteq \AA(f(v))\cap \AA(S)$ is a direct
      consequence of the fact that $f$ preserves labels and paths, while
      the converse inclusion is a direct consequence of the fact that $f$
      preserves labels and reflects paths.  \qed
    \end{proof}

    \begin{lemma}
    \label{lema-previ-tot}
    Let $f:S\to T$ be a weak topological embedding of $\AA$-trees.  Then:
    \begin{itemize}
    \item[(i)] $\LL(S)= \LL(T|\AA(S))$.
    \item[(ii)] $f$ induces a weak topological embedding $f:S\to
      T|\AA(S)$.
    \end{itemize}
    \end{lemma}

    \begin{proof}
      Notice first of all that $\AA(S)\subseteq \AA(T)$, because $f$
      preserves labels, and therefore it makes sense to define the
      restriction $T|\AA(S)$; actually, the nodes of $T$ with labels in
      $\AA(S)$ are exactly the images of the labeled nodes of $S$.  To
      simplify the notations, we shall denote in the rest of this proof
      $T|\AA(S)$ by $T'$.
      
      To prove (i), it is enough to check that the leaves of $T'$ are
      exactly the images of leaves of $S$ under $f$.  And recall that
      $w\in V(T')$ is a leaf of $T'$ if and only if $w=f(v_{S,A})$ for
      some $A\in \AA(S)$ and $\AA_{T}(w)\cap \AA(S)=\{A\}$.  Since, by the
      previous lemma, $\AA(f(v_{S,A}))\cap \AA(S)=\AA(v_{S,A})$, we deduce
      that $w\in V(T')$ is a leaf of $T'$ if and only if $w=f(v_{S,A})$
      for some $A\in \AA(S)$ such that $\AA(v_{S,A})=\{A\}$, that is, if
      and only if $w=f(v_{S,A})$ for some leaf $v_{S,A}$ of $S$, as we
      wanted to prove.
      
      As far as (ii) goes, let us prove first that $f(V(S))\subseteq
      V(T')$.  Let $v\in V(S)$.  If it is a leaf of $S$, then, as we have
      just seen, $f(v)\in V(T')$.  If $v$ is not a leaf of $S$, then there
      is a path in $S$ from $v$ to some leaf $v'$.  Since $f$ preserves
      paths, there is a path in $T$ from $f(v)$ to $f(v')$, and $f(v')$ is
      labeled in $\AA(S)$.  Therefore, by the definition of restriction of
      an $\AA$-tree, $f(v)\in V(T')$, too.
      
      This proves that $f(V(S))\subseteq V(T')$.  And then it is
      straightforward to deduce that $f:S\to T'$ is injective, preserves
      labels, and that it preserves and reflects paths, from the
      corresponding properties for $f:S\to T$.  \qed
    \end{proof}

    Now we can prove that, as we announced, weak topological embeddings
    capture ancestral displaying.

    \begin{proposition}
    \label{prop:disp}
    Let $S$ and $T$ be two $\AA$-trees, and let $S'$ be the semi-labeled
    tree obtained from $S$ by removing the elementary unlabeled nodes in
    it and replacing by arcs the maximal paths with all their intermediate
    nodes elementary and unlabeled.

    Then, $T$ ancestrally displays $S$ if and only if there exists a weak
    topological embedding $f:S'\to T$.
    \end{proposition}

    \begin{proof}
      Assume that $T$ ancestrally displays $S$, and in particular that
      $\AA(S)\subseteq \AA(T)$ and $\CC_{\AA}(S)\subseteq
      \CC_{\AA}(T|\AA(S))$; to simplify the notations, we shall denote
      $T|\AA(S)$ by $T''$.  Since elementary unlabeled nodes do not
      contribute any new member to the cluster representation,
      $\CC_{\AA}(S)=\CC_{\AA}(S')$.  Therefore, $\CC_{\AA}(S')\subseteq
      \CC_{\AA}(T'')$.
      
      We define the mapping
    $$
    \begin{array}{rrcl}
	f: & V(S') & \to & V(T'')\\
	& v & \mapsto & v_{T'',\AA(v)}
    \end{array}
    $$
    Let us check that this mapping defines a weak topological embedding
    $f:S'\to T''$.

    \begin{itemize}
    \item \emph{It is injective}. Let $v,w$ be two different nodes of
      $S'$.  Since every node in $S'$ is the most recent common ancestor
      of its labeled descendants, that is, $x=v_{S',\AA(x)}$ for every
      $x\in V(S')$, we have that $\AA(v)\neq \AA(w)$.  And then, since
      $\CC_{\AA}(S')\subseteq \CC_{\AA}(T'')$, it turns out that
      $\AA(v),\AA(w)$ are two different members of $\CC_{\AA}(T'')$, and
      hence $\AA(v_{T'',\AA(v)})=\AA(v)\neq \AA(w)=\AA(v_{T'',\AA(w)})$,
      which clearly implies that $v_{T'',\AA(v)}\neq v_{T'',\AA(w)}$.
      
    \item \emph{It preserves labels}.  Let $A\in \AA(S')$ and
      $v=v_{S',A}$.  Then, $f(v)=v_{T'', \AA(v_{S',A})}$ is labeled $A$
      because, by the second property of ancestral displaying, the labeled
      nodes in $S'$ that are descendants of $v$ are exactly the labeled
      nodes in $T''$ that are descendants of $v_{T'',A}$, and therefore
      $v_{T'',A}$ is the least common ancestor of the nodes with labels in
      $\AA(v_{S',A})$, that is, $v_{T'',A}=v_{T'', \AA(v_{S',A})}=f(v)$,
      as we claimed.
      
    \item \emph{It preserves and reflects paths}.  Since
      $\AA(v)=\AA(f(v))$ for every $v\in V(S')$, we have the following
      sequence of equivalences: for every $v,w\in V(S')$,
    $$
    \begin{array}{l}
    \mbox{there exists a non-trivial path $v\cami w$}\\
    \qquad\qquad\qquad\qquad \Longleftrightarrow
    \AA(w)\subsetneq \AA(v)\\
    \qquad\qquad\qquad\qquad \Longleftrightarrow
    \AA(f(w))\subsetneq \AA(f(v))\\
    \qquad\qquad \qquad\qquad\Longleftrightarrow
    \mbox{there exists a non-trivial path $f(v)\cami f(w)$.}
    \end{array}
    $$
    The implications $\Leftarrow$ in the first equivalence and
    $\Rightarrow$ in the last equivalence are given by
    Corollary~\ref{lem-int2}, while the converse implication in both cases
    is entailed by the fact that $v$, $w$, $f(v)$, and $f(w)$ are most recent common
    ancestors of sets of labeled nodes, and then non-trivial paths between
    them imply strict inclusions of sets of labels of descendants.
    \end{itemize}

    So, we have a weak topological embedding $f:S'\to T''$, and since
    $T''$ is a weak $\AA$-subtree of $T$, it induces a weak topological
    embedding $f:S'\to T$, as we wanted to prove.

    Conversely, assume that we have a weak topological embedding $f:S'\to
    T$.  Then:
    \begin{itemize}
    \item $\AA(S)=\AA(S')\subseteq \AA(T)$ because $f$ preserves labels.
    \item For every $A,B\in \AA(S)$, by construction, $v_{S,A}=v_{S',A}$
      and $v_{S,B}=v_{S',B}$, and there exists a path $v_{S,A}\cami
      v_{S,B}$ in $S$ if and only if there exists a path $v_{S',A}\cami
      v_{S',B}$ in $S'$.  Moreover, since $f$ preserves labels and
      preserves and reflects paths, there exists a path $v_{S',A}\cami
      v_{S',B}$ in $S'$ if and only if there exists a path
      $v_{T,A}=f(v_{S',A})\cami f(v_{S',B})=v_{T,B}$ in $T$.  Combining
      these equivalences, we obtain that, for every $A,B\in \AA(S)$, there
      exists a path $v_{S,A}\cami v_{S,B}$ in $S$ if and only if there
      exists a path $v_{T,A}\cami v_{T,B}$ in $T$.
    \item Let $X\in \CC_{\AA}(S)$ and let $v=v_{S,X}=v_{S',X}$.  It turns
      out that $\AA_{T|\AA(S)}(f(v))=X$.  Indeed, by
      Lemma~\ref{lema-previ-tot}, $f:S'\to T$ induces a weak topological
      embedding $f:S'\to T|\AA(S')=T|\AA(S)$ and then, by
      Lemma~\ref{Av=Af(v)},
      $\AA_{T|\AA(S)}(f(v))=\AA_{S'}(v)=\AA_{S}(v)=X$.
      
      Therefore, $X\in \CC(T|\AA(S))$, and, being $X$ arbitrary, we
      conclude that $\CC_{\AA}(S)\subseteq \CC(T|\AA(S))$.
    \end{itemize}
    This proves that $T$ ancestrally displays $S$.\qed
    \end{proof}

    Now, recall from \cite{semple.ea:2004} the notion of ancestral
    compatibility.

    \begin{definition}
    \label{def:ac}
    Two $\AA$-trees $T_{1},T_{2}$ are \emph{ancestrally compatible} when
    there exists an $\AA$-tree that ancestrally displays both of them.  If
    two $\AA$-trees are not ancestrally compatible, we say that they are
    \emph{ancestrally incompatible}.
    \end{definition}

    Weak topological embeddings have been defined as they have so
    ancestral compatibility turns out to be exactly the same as
    `compatibility for weak topological embeddings.'

    \begin{proposition}
      Two $\AA$-trees $T_{1},T_{2}$ are ancestrally compatible if and only
      if they have a common weak $\AA$-supertree, that is, if and only if
      they admit a weak topological embedding into a same $\AA$-tree.
    \end{proposition}

    \begin{proof}
      For every $\ell=1,2$, let $T_{\ell}'$ be the semi-labeled tree
      obtained by removing the elementary unlabeled nodes in $T_{\ell}$
      and replacing by arcs the maximal paths with all their intermediate
      nodes elementary and unlabeled.
      
      Assume that there exist weak topological embeddings $f_{1}:T_{1}\to
      T$ and $f_{2}:T_{2}\to T$ of $T_{1}$ and $T_{2}$ into a same
      $\AA$-tree $T$.  Since each $T_{\ell}'$ is a weak $\AA$-subtree of
      the corresponding $T_{\ell}$, each one of these weak topological
      embeddings induces a weak topological embedding
      $f_{\ell}':T_{\ell}'\to T$, showing that $T$ ancestrally displays
      $T_{1}$ and $T_{2}$.
      
      Conversely, assume that there exist weak topological embeddings
      $g_{1}:T_{1}'\to T$ and $g_{2}:T_{2}'\to T$ of $T_{1}'$ and $T_{2}'$
      into a same $\AA$-tree $T$.  Let $\tilde{T}$ be the $\AA$-tree
      obtained from $T$ in the following way.  For every arc $(v,w)\in
      E(T)$, if there exists an arc $(v_{\ell},w_{\ell})$ in one
      $T_{\ell}$ such that $g_{\ell}(v_{\ell})=v$ and
      $g_{\ell}(w_{\ell})=w$, we split the arc $(v,w)$ in $T$ into a path
      $v\cami w$, with all its intermediate nodes elementary and
      unlabeled, of length equal to the length of the path $v_{\ell}\cami
      w_{\ell}$; if there are arcs $(v_{1},w_{1})\in E(T_{1})$ and
      $(v_{2},w_{2})\in E(T_{2})$ such that $g_{1}(v_{1})=g_{2}(v_{2})=v$
      and $g_{1}(w_{1})=g_{2}(w_{2})=w$, then we split the arc $(v,w)$ in
      $T$ into a path $v\cami w$ as before, but now of length the maximum
      of the lengths of the paths $v_{1}\cami w_{1}$ and $v_{2}\cami
      w_{2}$.  It is clear then that each $g_{T}:T\to T_{0}$ can be
      extended to a weak topological embedding $\tilde{g}_{T}:T\to
      \tilde{T}$.  \qed
    \end{proof}

    From now on, we shall use this characterization of ancestral
    compatibility as the working definition of it.

    The main result of this paper will establish that ancestral
    compatibility is equivalent to local compatibility.  To prove it, we
    shall need a preliminary result, Proposition~\ref{lem-restr}, which
    establishes that ancestral compatibility of two $\AA$-trees can be
    checked at the level of $\bar{T}_{1}$ and $\bar{T_{2}}$, as it was
    also the case for local compatibility.

    \begin{lemma}
    \label{cor-presuper}
    \label{cor:wtehat}
    Let $T_{1}$ and $T_{2}$ be two $\AA$-trees and let $\bar{T}_{1}$ and
    $\bar{T}_{2}$ be their $\AA$-subtrees described in
    Definition~\ref{def:hat}.  If $T_{1}$ and $T_{2}$ are ancestrally
    com\-pat\-i\-ble, then $\LL(\bar{T}_{1})=\LL(\bar{T}_{2})$.
    \end{lemma}

    \begin{proof}
      Assume that $T_{1}$ and $T_{2}$ are ancestrally com\-pat\-i\-ble.
      Then, since $\bar{T}_{1}$ and $\bar{T}_{2}$ are weak $\AA$-subtrees
      of $T_{1}$ and $T_{2}$, respectively, it is clear that they are also
      ancestrally com\-pat\-i\-ble; let $f_{1}:\bar{T}_{1}\to T$ and
      $f_{2}:\bar{T}_{2}\to T$ be weak topological embeddings.  Recall
      that $\AA(\bar{T}_{1})=\AA(\bar{T}_{2})$.
      
      If $A\in \LL(\bar{T}_{1})$, then
      $\AA_{\bar{T}_{1}}(v_{\bar{T}_{1},A})=\{A\}$ and hence
    $$
    \begin{array}{rl}
    \AA_{\bar{T}_{2}}(v_{\bar{T}_{2},A}) & =\AA_{T}(f_{2}(v_{\bar{T}_{2},A}))\cap \AA(\bar{T}_{2})=
    \AA_{T}(v_{T,A})\cap \AA(\bar{T}_{2})\\ & =
    \AA_{T}(f_{1}(v_{\bar{T}_{1},A}))\cap \AA(\bar{T}_{1})=
    \AA_{\bar{T}_{1}}(v_{\bar{T}_{1},A})=
    \{A\},
    \end{array}
    $$
    which says that $v_{\bar{T}_{2},A}$ is a leaf of $\bar{T}_{2}$ and
    thus $A\in \LL(\bar{T}_{2})$.

    This proves that $\LL(\bar{T}_{1})\subseteq \LL(\bar{T}_{2})$ and, by
    symmetry, the equality between these two sets.  \qed
    \end{proof}

    \begin{proposition}
    \label{lem-restr}
    Let $T_{1}$ and $T_{2}$ be $\AA$-trees and let $\bar{T}_{1}$ and
    $\bar{T}_{2}$ be their $\AA$-subtrees described in
    Definition~\ref{def:hat}.  Then, $T_{1}$ and $T_{2}$ are ancestrally
    com\-pat\-i\-ble if and only if $\bar{T}_{1}$ and $\bar{T}_{2}$ are
    ancestrally com\-pat\-i\-ble.
    \end{proposition}

    \begin{proof}
      As we have seen in the proof of the last lemma, if $T_{1}$ and
      $T_{2}$ are ancestrally com\-pat\-i\-ble, then $\bar{T}_{1}$ and
      $\bar{T}_{2}$ are also so.  Conversely, let $f_{1}:\bar{T}_{1}\to T$
      and $f_{2}:\bar{T}_{2}\to T$ be two weak topological embeddings.  By
      the last lemma, we know that $\LL(\bar{T}_{1})=\LL(\bar{T}_{2})$.
      Recall, moreover, that
      $\AA(\bar{T}_{1})=\AA(\bar{T}_{2})=\AA(T_{1})\cap \AA(T_{2})$.
      
      By Lemma~\ref{lema-previ-tot}, $f_{1}$ and $f_{2}$ induce weak
      topological embeddings into the restriction of $T$ to
      $\AA(\bar{T}_{1})=\AA(\bar{T}_{2})$.  Therefore, by replacing $T$ by
      this $\AA$-subtree if necessary, we shall assume without any loss of
      generality that $\LL(T)=\LL(\bar{T}_{1})=\LL(\bar{T}_{2})$.  We
      shall also assume, again without any loss of generality, that
      $\AA(T)=\AA(\bar{T}_{1})=\AA(\bar{T}_{2})$: we simply remove from
      $T$ the labels that do not belong to this set.
      
      Finally, we shall assume that there does not exist any pair of
      different labels $A_{1},A_{2}$ such that $v_{T_{1},A_{1}}\in
      V(\bar{T}_{1})$ and $v_{T_{2},A_{2}}\in V(\bar{T}_{2})$ and
      $f_{1}(v_{T_{1},A_{1}})=f_{2}(v_{T_{2},A_{2}})$.  Indeed, assume
      that such a pair of labels exists.  Then, to begin with,
      $A_{1},A_{2}\notin \AA(T_{1})\cap \AA(T_{2})$: if, say, $A_{2}\in
      \AA(T_{1})\cap \AA(T_{2})$ then, since $f_{1}$ and $f_{2}$ preserve
      labels, it happens that
      $f_{2}(v_{T_{2},A_{2}})=f_{1}(v_{T_{1},A_{2}})$ and then
      $v_{T_{1},A_{2}}=v_{T_{1},A_{1}}$, that is, $A_{2}=A_{1}$.
      Therefore, $v_{T_{1},A_{1}}$ and $v_{T_{2},A_{2}}$ do not keep their
      labels in $\bar{T}_{1}$ and $\bar{T}_{2}$.  Now, given the node
      $w=f_{1}(v_{T_{1},A_{1}})=f_{2}(v_{T_{2},A_{2}})$ (which, by what we
      have just discussed, will be unlabeled, either), we `blow out' it by
      adding a new node $w'$, splitting the arc going from $w$'s parent
      $w_{0}$ to $w$ into two arcs $(w_{0},w'), (w',w)$ ---if $w$ was the
      root of $T$, we simply add a new arc $(w',w)$--- and redefining
      $f_{1}$ by sending $v_{T_{1},A_{1}}$ to $w'$ while we do not change
      $f_{2}$ (alternatively, we could have redefined $f_{2}$, by sending
      $v_{T_{2},A_{2}}$ to $w'$, and left $f_{1}$ unchanged).  It is
      straightforward to check that the new mapping $f_{1}$ obtained in
      this way and the `old' $f_{2}$ are still weak topological embeddings
      from $T_{1}$ and $T_{2}$ to the new $\AA$-tree.  After repeating
      this process as many times as necessary, and still calling $T$ the
      target $\AA$-tree obtained at the end, we obtain weak topological
      embeddings $f_{1}:\bar{T}_{1}\to T$ and $f_{2}:\bar{T}_{2}\to T$ as
      we assumed at the beginning of this paragraph.
      
      We shall expand this common weak $\AA$-supertree $T$ of
      $\bar{T}_{1}$ and $\bar{T}_{2}$ to a common weak $\AA$-supertree of
      $T_{1}$ and $T_{2}$.  To begin with, we expand $T$ to an
      $\AA$-labeled graph $T'$ by ``adding $T_{1}- \bar{T}_{1}$'' to it.
      More specifically, to obtain $T'$, we add to $T$ all nodes in
      $V(T_{1})- V(\bar{T}_{1})$, and arcs of two types: on the one hand,
      those between these nodes in $T_{1}$, and on the other hand, for
      every arc $(a,b)\in E(T_{1})$ with $a\in V(\bar{T}_{1})$ and $b\in
      V(T_{1})- V(\bar{T}_{1})$, an arc between $f_{1}(a)$ and $b$ in
      $T'$.  As far as the labels go, on the one hand the nodes of $T'$
      belonging to $V(T_{1})- V(\bar{T}_{1})$ inherit their labels, and on
      the other hand the nodes in $T'$ that are images of nodes in
      $\bar{T}_{1}$ labeled in $\AA(T_{1})- \AA(\bar{T}_{1})$, are labeled
      with this label.  None of the labels we add in this way could be
      present in $T$, because otherwise they would have belonged to
      $\AA(\bar{T}_{1})$, which is impossible, and no already labeled node
      in $T$ receives a second label, because the nodes labeled in $T$
      received their labels from $\bar{T}_{1}$.
      
      This $T'$ is clearly an $\AA$-tree, and has $T$ as a weak
      $\AA$-subtree: actually, $T=T'|\LL(T)$.  Therefore, it is a weak
      $\AA$-supertree of $\bar{T}_{2}$.  And it is also a weak
      $\AA$-supertree of $T_{1}$.  Indeed, consider the mapping
      $f'_{1}:V(T_{1})\to V(T')$ that is defined on $V(\bar{T}_{1})$ as
      the original embedding $f_{1}:V(\bar{T}_{1})\to V(T)$ and on
      $V(T_{1}) - V(\bar{T}_{1})$ as the identity.  It is clearly
      injective and preserves labels.  Moreover, it preserves paths,
      because $f_{1}$ sends arcs in $\bar{T}_{1}$ to paths in $T$, and
      arcs outside $\bar{T}_{1}$ become arcs in $T'$; and it reflects
      paths, because it reflects paths in $T$ and the arcs that have been
      added come from arcs in $T_{1}$.
      
      So, $T'$ is a common weak $\AA$-supertree of $T_{1}$ and
      $\bar{T}_{2}$.  Now, we expand $T'$ to a new $\AA$-tree $T''$ by
      means of a similar process, but now ``adding $T_{2}- \bar{T}_{2}$''
      to it.  We add to $T'$ all nodes in $V(T_{2})- V(\bar{T}_{2})$, all
      arcs between these nodes in $T_{2}$, an arc $(f_{2}(a),b)$ for every
      arc $(a,b)\in E(T_{2})$ with $a\in V(\bar{T}_{2})$ and $b\in
      V(T_{2})- V(\bar{T}_{2})$.  The new nodes, coming from $V(T_{2})-
      V(\bar{T}_{2})$, are labeled as they were in $T_{2}$, while the old
      ones receive their labels from $T_{2}$, if any and necessary.  No
      new label added in this way could be already present in $T'$. And no
      already labeled node receives a second label, because the images of
      $f'_{1}:T_{1}\to T'$ and $f_{2}:\bar{T}_{2}\to T'$ are still
      disjoint except for the nodes with labels in
      $\AA(T_{1})\cap\AA(T_{2})$.
      
      The $\AA$-labeled graph $T''$ obtained in this way is again an
      $\AA$-tree, and now it is a weak $\AA$-supertree of $T_{1}$ and of
      $T_{2}$: the proof is similar to the previous one in the case of
      $T'$.  Therefore, $T_{1}$ and $T_{2}$ are ancestrally
      com\-pat\-i\-ble, as we wanted to prove.  \qed
    \end{proof}

    \begin{example}
    \label{ex:finsec4}
    Consider the semi-labeled trees $T_{1}$ and $T_{2}$ described in
    Fig.~\ref{fig:exfinsec4-1}.  The corresponding $\AA$-trees
    $\bar{T}_{1}$ and $\bar{T}_{2}$, which are no longer semi-labeled
    trees, are described in Fig.~\ref{fig:exfinsec4-2}; notice that the
    nodes $c$, $h$ and $i$ are no longer labeled in these trees.
      
    The $\AA$-trees $\bar{T}_{1}$ and $\bar{T}_{2}$ are ancestrally
    compatible.  A weak common $\AA$-super\-tree of them is given by the
    $\AA$-tree $T$ described in Fig.~\ref{fig:exfinsec4-3.5}, together
    with the weak topological embeddings $f_{1}:\bar{T}_{1}\to T$ and
    $f_{2}:\bar{T}_{2}\to T$ that are indicated by assigning in the
    picture to each non-labeled node in $T$ its preimages under $f_{1}$
    and $f_{2}$.  Notice that $\AA(T)=\AA(\bar{T}_{1})=\AA(\bar{T}_{2})$,
    but $f_{1}(v_{T_{1},C})= f_{2}(v_{T_{2},H})$.  To avoid it, we blow up
    this node into an arc and we separate these two images: the
    corresponding new weak $\AA$-supertree $T$ is described in
    Fig.~\ref{fig:exfinsec4-3}.  Now, the new weak topological embeddings
    $f_{1}$ and $f_{2}$ satisfy the assumptions in the proof of the last
    proposition.
      
    The $\AA$-trees $T'$ and $T''$ that are successively obtained by first
    `adding $T_{1}-\bar{T}_{1}$ to $T$' and then `adding
    $T_{2}-\bar{T}_{2}$ to $T'$' are described in
    Figs.~\ref{fig:exfinsec4-4} and~\ref{fig:exfinsec4-5}, respectively.
    At the end, $T''$ is a weak common $\AA$-supertree of $T_{1}$ and
    $T_{2}$ under the embeddings indicated as before.

    \begin{figure}[htb]
\begin{center}
        \begin{picture}(290,75)(-10,-10)
        \put(0,0){\line(1,1){60}}
        \put(40,0){\line(-1,1){20}} 
        \put(60,0){\line(-1,2){20}} 
        \put(120,0){\line(-1,1){60}} 
        \put(80,0){\line(1,1){20}} 
        \put(100,0){\line(0,1){20}} 
        \put(0,0){\circle*{3}} 
        \put(20,20){\circle*{3}} 
        \put(40,40){\circle*{3}} 
        \put(40,0){\circle*{3}} 
        \put(60,0){\circle*{3}} 
        \put(60,60){\circle*{3}} 
        \put(100,20){\circle*{3}} 
        \put(120,0){\circle*{3}} 
        \put(100,0){\circle*{3}} 
        \put(80,0){\circle*{3}} 
        \put(120,0){\circle*{3}} 
        \put(0,-5){\makebox(0,0)[t]{$A$}}
        \put(40,-5){\makebox(0,0)[t]{$B$}}
        \put(60,-5){\makebox(0,0)[t]{$D$}}
        \put(80,-5){\makebox(0,0)[t]{$E$}}
        \put(100,-5){\makebox(0,0)[t]{$F$}}
        \put(120,-5){\makebox(0,0)[t]{$G$}}
        \put(15,20){\makebox(0,0)[r]{$C$}}
        \put(35,40){\makebox(0,0)[r]{$1$}}
        \put(55,60){\makebox(0,0)[r]{$2$}}
        \put(105,20){\makebox(0,0)[l]{$3$}}
        \put(100,50){\makebox(0,0)[l]{$T_{1}$}}
        \put(150,0){\line(1,1){60}}
        \put(190,0){\line(-1,1){20}} 
        \put(210,0){\line(0,1){60}} 
        \put(270,0){\line(-1,1){60}} 
        \put(230,0){\line(0,1){40}} 
        \put(150,0){\circle*{3}} 
        \put(190,0){\circle*{3}} 
        \put(230,0){\circle*{3}} 
        \put(210,0){\circle*{3}} 
        \put(270,0){\circle*{3}} 
        \put(170,20){\circle*{3}} 
        \put(210,60){\circle*{3}} 
        \put(230,40){\circle*{3}} 
        \put(250,20){\circle*{3}} 
        %
        \put(150,-5){\makebox(0,0)[t]{$A$}}
        \put(190,-5){\makebox(0,0)[t]{$B$}}
        \put(210,-5){\makebox(0,0)[t]{$E$}}
        \put(230,-5){\makebox(0,0)[t]{$J$}}
        \put(270,-5){\makebox(0,0)[t]{$K$}}
        \put(165,20){\makebox(0,0)[r]{$H$}}
        \put(205,60){\makebox(0,0)[r]{$4$}}
        \put(235,40){\makebox(0,0)[l]{$I$}}
        \put(255,20){\makebox(0,0)[l]{$G$}}
        \put(250,50){\makebox(0,0)[l]{$T_{2}$}}
        \end{picture}
        \caption{\label{fig:exfinsec4-1}%
          The semi-labeled trees $T_{1},T_{2}$ in Example~\ref{ex:finsec4}}
\end{center}
    \end{figure}
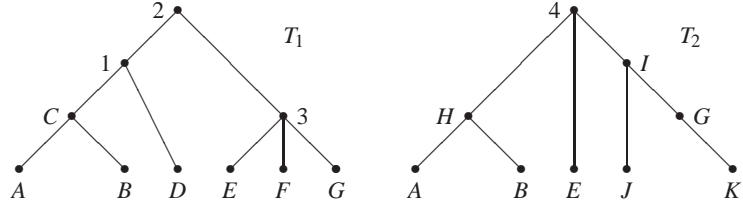

    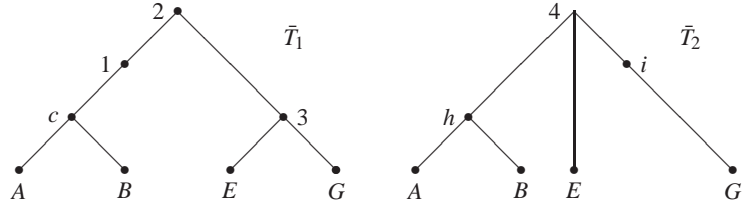
\begin{figure}[htb]
\begin{center}
        \begin{picture}(290,75)(-10,-10)
        \put(0,0){\line(1,1){60}}
        \put(40,0){\line(-1,1){20}} 
        \put(120,0){\line(-1,1){60}} 
        \put(80,0){\line(1,1){20}} 
        \put(0,0){\circle*{3}} 
        \put(20,20){\circle*{3}} 
        \put(40,40){\circle*{3}} 
        \put(40,0){\circle*{3}} 
        \put(60,60){\circle*{3}} 
        \put(100,20){\circle*{3}} 
        \put(120,0){\circle*{3}} 
        \put(80,0){\circle*{3}} 
        \put(120,0){\circle*{3}} 
        \put(0,-5){\makebox(0,0)[t]{$A$}}
        \put(40,-5){\makebox(0,0)[t]{$B$}}
        \put(80,-5){\makebox(0,0)[t]{$E$}}
        \put(120,-5){\makebox(0,0)[t]{$G$}}
        \put(15,20){\makebox(0,0)[r]{$c$}}
        \put(35,40){\makebox(0,0)[r]{$1$}}
        \put(55,60){\makebox(0,0)[r]{$2$}}
        \put(105,20){\makebox(0,0)[l]{$3$}}
        \put(100,50){\makebox(0,0)[l]{$\bar{T}_{1}$}}
        \put(150,0){\line(1,1){60}}
        \put(190,0){\line(-1,1){20}} 
        \put(210,0){\line(0,1){60}} 
        \put(270,0){\line(-1,1){60}} 
        \put(150,0){\circle*{3}} 
        \put(190,0){\circle*{3}} 
        \put(210,0){\circle*{3}} 
        \put(270,0){\circle*{3}} 
        \put(170,20){\circle*{3}} 
        \put(230,40){\circle*{3}} 
        %
        \put(150,-5){\makebox(0,0)[t]{$A$}}
        \put(190,-5){\makebox(0,0)[t]{$B$}}
        \put(210,-5){\makebox(0,0)[t]{$E$}}
        \put(270,-5){\makebox(0,0)[t]{$G$}}
        \put(165,20){\makebox(0,0)[r]{$h$}}
        \put(205,60){\makebox(0,0)[r]{$4$}}
        \put(235,40){\makebox(0,0)[l]{$i$}}
        \put(250,50){\makebox(0,0)[l]{$\bar{T}_{2}$}}
        \end{picture}
        \caption{\label{fig:exfinsec4-2}%
          The $\AA$-trees $\bar{T}_{1},\bar{T}_{2}$ corresponding to the
          semi-labeled trees $T_{1},T_{2}$ in Fig.~\ref{fig:exfinsec4-1}}
\end{center}
    \end{figure}

    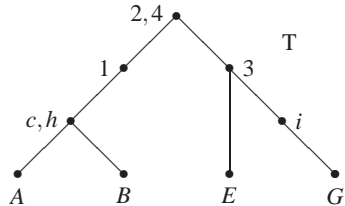
\begin{figure}[htb]
\begin{center}
        \begin{picture}(110,75)(-10,-10)
        \put(0,0){\line(1,1){60}}
        \put(40,0){\line(-1,1){20}} 
        \put(120,0){\line(-1,1){60}} 
        \put(80,0){\line(0,1){40}} 
        \put(0,0){\circle*{3}} 
        \put(20,20){\circle*{3}} 
        \put(40,40){\circle*{3}} 
        \put(40,0){\circle*{3}} 
        \put(60,60){\circle*{3}} 
        \put(80,40){\circle*{3}} 
        \put(100,20){\circle*{3}} 
        \put(120,0){\circle*{3}} 
        \put(80,0){\circle*{3}} 
        \put(120,0){\circle*{3}} 
        \put(0,-5){\makebox(0,0)[t]{$A$}}
        \put(40,-5){\makebox(0,0)[t]{$B$}}
        \put(80,-5){\makebox(0,0)[t]{$E$}}
        \put(120,-5){\makebox(0,0)[t]{$G$}}
        \put(15,20){\makebox(0,0)[r]{$c,h$}}
        \put(35,40){\makebox(0,0)[r]{$1$}}
        \put(55,60){\makebox(0,0)[r]{$2,4$}}
        \put(105,20){\makebox(0,0)[l]{$i$}}
        \put(85,40){\makebox(0,0)[l]{$3$}}
        \put(100,50){\makebox(0,0)[l]{$$T$$}}
        \end{picture}
        \caption{\label{fig:exfinsec4-3.5}%
          A weak common $\AA$-supertree of $\bar{T}_{1}$ and $\bar{T}_{2}$}
\end{center}
    \end{figure}

    \begin{figure}[htb]
\begin{center}
        \begin{picture}(180,95)(-10,-10)
        \put(0,0){\line(1,1){80}}
        \put(40,0){\line(-1,1){20}} 
        \put(160,0){\line(-1,1){80}} 
        \put(80,0){\line(1,3){20}} 
        \put(0,0){\circle*{3}} 
        \put(20,20){\circle*{3}} 
        \put(40,40){\circle*{3}} 
        \put(60,60){\circle*{3}} 
        \put(80,80){\circle*{3}} 
        \put(40,0){\circle*{3}} 
        \put(80,0){\circle*{3}} 
        \put(100,60){\circle*{3}} 
        \put(130,30){\circle*{3}} 
        \put(160,0){\circle*{3}} 
        \put(0,-5){\makebox(0,0)[t]{$A$}}
        \put(40,-5){\makebox(0,0)[t]{$B$}}
        \put(80,-5){\makebox(0,0)[t]{$E$}}
        \put(160,-5){\makebox(0,0)[t]{$G$}}
        \put(15,20){\makebox(0,0)[r]{$h$}}
        \put(35,40){\makebox(0,0)[r]{$c$}}
        \put(55,60){\makebox(0,0)[r]{$1$}}
        \put(75,80){\makebox(0,0)[r]{$2,4$}}
        \put(105,60){\makebox(0,0)[l]{$3$}}
        \put(135,30){\makebox(0,0)[l]{$i$}}
        \put(125,70){\makebox(0,0)[l]{$T$}}
        \end{picture}
        \caption{\label{fig:exfinsec4-3}%
          The new $\AA$-tree $T$ obtained after blowing out the node $c,h$ in
          the $\AA$-tree $T$ in Fig.~\ref{fig:exfinsec4-3.5}}
\end{center}
    \end{figure}
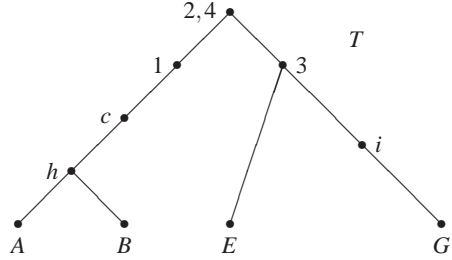

    \begin{figure}[htb]
\begin{center}
        \begin{picture}(180,95)(-10,-10)
        \put(0,0){\line(1,1){80}}
        \put(40,0){\line(-1,1){20}} 
        \put(60,0){\line(0,1){60}}
        \put(100,0){\line(0,1){60}}
        \put(160,0){\line(-1,1){80}} 
        \put(80,0){\line(1,3){20}} 
        \put(0,0){\circle*{3}} 
        \put(20,20){\circle*{3}} 
        \put(40,40){\circle*{3}} 
        \put(60,60){\circle*{3}} 
        \put(80,80){\circle*{3}} 
        \put(40,0){\circle*{3}} 
        \put(60,0){\circle*{3}} 
        \put(80,0){\circle*{3}} 
        \put(100,0){\circle*{3}} 
        \put(100,60){\circle*{3}} 
        \put(130,30){\circle*{3}} 
        \put(160,0){\circle*{3}} 
        \put(0,-5){\makebox(0,0)[t]{$A$}}
        \put(40,-5){\makebox(0,0)[t]{$B$}}
        \put(60,-5){\makebox(0,0)[t]{$D$}}
        \put(80,-5){\makebox(0,0)[t]{$E$}}
        \put(100,-5){\makebox(0,0)[t]{$F$}}
        \put(160,-5){\makebox(0,0)[t]{$G$}}
        \put(15,20){\makebox(0,0)[r]{$h$}}
        \put(35,40){\makebox(0,0)[r]{$C$}}
        \put(55,60){\makebox(0,0)[r]{$1$}}
        \put(80,85){\makebox(0,0)[b]{$2,4$}}
        \put(105,60){\makebox(0,0)[l]{$3$}}
        \put(135,30){\makebox(0,0)[l]{$i$}}
        \put(125,70){\makebox(0,0)[l]{$T'$}}
        \end{picture}
        \caption{\label{fig:exfinsec4-4}%
          The $\AA$-tree $T'$ obtained by `adding $T_{1}-\bar{T}_{1}$' to $T$}
\end{center}
    \end{figure}
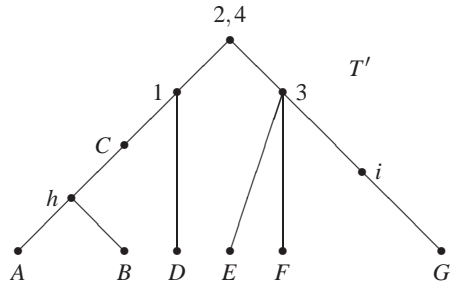

    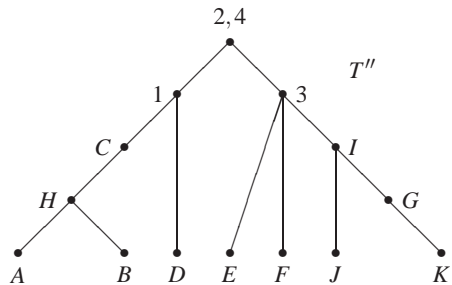
\begin{figure}[htb]
\begin{center}
        \begin{picture}(180,95)(-10,-10)
        \put(0,0){\line(1,1){80}}
        \put(40,0){\line(-1,1){20}} 
        \put(60,0){\line(0,1){60}}
        \put(100,0){\line(0,1){60}}
        \put(120,0){\line(0,1){40}}
        \put(160,0){\line(-1,1){80}} 
        \put(80,0){\line(1,3){20}} 
        \put(0,0){\circle*{3}} 
        \put(20,20){\circle*{3}} 
        \put(40,40){\circle*{3}} 
        \put(60,60){\circle*{3}} 
        \put(80,80){\circle*{3}} 
        \put(40,0){\circle*{3}} 
        \put(60,0){\circle*{3}} 
        \put(80,0){\circle*{3}} 
        \put(100,0){\circle*{3}} 
        \put(120,0){\circle*{3}} 
        \put(160,0){\circle*{3}} 
        \put(100,60){\circle*{3}} 
        \put(120,40){\circle*{3}} 
        \put(140,20){\circle*{3}} 
        \put(0,-5){\makebox(0,0)[t]{$A$}}
        \put(40,-5){\makebox(0,0)[t]{$B$}}
        \put(60,-5){\makebox(0,0)[t]{$D$}}
        \put(80,-5){\makebox(0,0)[t]{$E$}}
        \put(100,-5){\makebox(0,0)[t]{$F$}}
        \put(120,-5){\makebox(0,0)[t]{$J$}}
        \put(160,-5){\makebox(0,0)[t]{$K$}}
        \put(145,20){\makebox(0,0)[l]{$G$}}
        \put(15,20){\makebox(0,0)[r]{$H$}}
        \put(35,40){\makebox(0,0)[r]{$C$}}
        \put(55,60){\makebox(0,0)[r]{$1$}}
        \put(80,85){\makebox(0,0)[b]{$2,4$}}
        \put(105,60){\makebox(0,0)[l]{$3$}}
        \put(125,40){\makebox(0,0)[l]{$I$}}
        \put(125,70){\makebox(0,0)[l]{$T''$}}
        \end{picture}
        \caption{\label{fig:exfinsec4-5}%
          The weak common $\AA$-supertree $T''$ of $T_{1}$ and $T_{2}$
          obtained by `adding $T_{2}-\bar{T}_{2}$' to $T'$}
\end{center}
    \end{figure}
    \end{example}

    \section{Main results}
    \label{sect:main}

    In this section we establish that local compatibility is the same as
    ancestral compatibility.  We also provide a characterization of the
    ancestral, or local, compatibility of a family of $\AA$-trees in terms
    of joint properties of their cluster representations.

    \begin{definition}\label{T12}
      Let $T_{1}$ and $T_{2}$ be two $\AA$-trees.
      \begin{enumerate}
      \item [(a)] Assume that $\AA(T_{1})=\AA(T_{2})$.  In this case, the
	\emph{join} of $T_{1}$ and $T_{2}$ is the $\AA$-labeled graph
	$T_{1,2}$ defined as follows.
	
	For every $\ell=1,2$ and for every $Y\in \CC_{\AA}(T_{\ell})$, let
	$$
	m_{\ell,Y}=\#\{v\in V(T_{\ell})\mid \AA_{T_{\ell}}(v)=Y\}.
	$$
	Set $\CC=\CC_{\AA}(T_{1})\cup \CC_{\AA}(T_{2})$.  Then:
	\begin{itemize}
	\item Its nodes are
	$$
	w_{Y,j} \quad \mbox{with $Y\in\CC$ and $j=1,\ldots,n_{Y}$,}
	$$
	where
	$n_{Y}=\max\{m_{1,Y},m_{2,Y}\}$.
	
      \item Its arcs are:
	$$
	\begin{array}{ll}
	(w_{Y,j},w_{Y,j-1})&\quad j=2,\ldots,n_{Y}\\
	(w_{Y,1},w_{Z,n_{Z}}) & \quad \mbox{if $Z\subsetneq Y$ and there is
	no $Z'\in\CC$ such that $Z\subsetneq Z'\subsetneq Y$.}
	\end{array}
	$$
	
      \item If there exists some $Y\in \CC$ such that
	$$
	Y=(\bigcup \{Z\in \CC\mid Z\subsetneq Y\})\sqcup\{A\}
	$$
	for some label $A\in \AA$, then the node $w_{Y,1}$ is labeled
	with this $A$.  In particular, the nodes $w_{A,1}$, with $\{A\}$
	any singleton in $\CC$, are labeled with the corresponding label
	$A$.
	\end{itemize}
	
	Now, for every $\ell=1,2$, we define a mapping
	$f_{\ell}:V(T_{\ell})\to V(T_{1,2})$ as follows.  For every $Y\in
	\CC_{\AA}(T_{\ell})$, let
	$\{x^{(\ell)}_{Y,1},\ldots,x^{(\ell)}_{Y,m_{\ell,Y}}\}\in
	V(T_{\ell})$ be the set of nodes of $T_{\ell} $ with cluster $Y$,
	ordered as follows: $x^{(\ell)}_{Y,1}=v_{T_{\ell},Y}$, and
	$(x^{(\ell)}_{Y,i+1},x^{(\ell)}_{Y,i})\in E(T_{\ell})$ for every
	$i=1,\ldots,m_{\ell,Y}-1$.
	
	With these notations, $f_{\ell}:V(T_{\ell}) \to V(T)$ is defined
	by
	\begin{center}
	  $f_{\ell}(x^{(\ell)}_{Y,i})=w_{Y,i}$ for every $Y\in \CC_{\AA}(T_{\ell})$
	  and $i=1,\ldots,m_{Y}$.
	\end{center}
	Since $\CC_{\AA}(T_{\ell})\subseteq \CC$ and, for every $Y\in
	\CC_{\AA}(T_{\ell})$, $m_{\ell,Y}\leq n_{Y}$, it is clear that
	$f_{\ell}$ is well defined and injective.
	
      \item [(b)] If $\AA(T_{1})\neq \AA(T_{2})$, let $\bar{T}_{1}$ and
	$\bar{T}_{2}$ be the $\AA$-subtrees of $T_{1}$ and $T_{2}$
	described in Definition~\ref{def:hat}.  Then, the \emph{join}
	$T_{1,2}$ of $T_{1}$ and $T_{2}$ is the result of applying the
	construction in the proof of Proposition~\ref{lem-restr} to the
	join $\bar{T}_{1,2}$ of $\bar{T}_{1}$ and $\bar{T}_{2}$ (that is,
	first blowing out into arcs the nodes that are images of pairs of
	nodes labeled with different labels, next `adding
	$T_{1}-\bar{T_{1}}$' to this $\AA$-tree, and finally `adding
	$T_{2}-\bar{T_{2}}$' to the result), and the mappings
	$f_{\ell}:V(T_{\ell})\to V({T}_{1,2})$, $\ell=1,2$, are obtained
	by extending the mappings $f_{\ell}:V(\bar{T}_{\ell})\to
	V(\bar{T}_{1,2})$ also in the way described in that proof.
    \end{enumerate}
    \end{definition}

    Notice that, by construction, the mappings $f_{l}:V(T_{l})\to
    V({T}_{1,2})$, $l=1,2$, are \emph{jointly surjective}, that is, every
    node of $T_{1,2}$ belongs to the image of one or the other.

    \begin{theorem}
    \label{th:ancestrally comp}
    Let $T_{1}$ and $T_{2}$ be two $\AA$-trees with
    $\AA(T_{1})=\AA(T_{2})$.  Then, the following assertions are
    equivalent:
    \begin{itemize}
    \item[(i)] $T_{1}$ and $T_{2}$ are ancestrally compatible.
    \item[(ii)] $T_{1}$ and $T_{2}$ are locally compatible.
    \item[(iii)] $\CC_{\AA}(T_{1})$ and $\CC_{\AA}(T_{2})$ satisfy jointly
      the following two conditions:
    \begin{itemize}
    \item For every $A\in \AA(T_{1})=\AA(T_{2})$, the smallest member of
      $\CC_{\AA}(T_{1})$ containing $A$ is equal to the smallest member of
      $\CC_{\AA}(T_{2})$ containing this label.
    \item For every $X\in \CC_{\AA}(T_{1})$ and $Y\in \CC_{\AA}(T_{2})$,
      if $X\cap Y\neq \emptyset$, then $X\subseteq Y$ or $Y\subseteq X$.
    \end{itemize}
    \item[(iv)] The join $T_{1,2}$ of $T_{1}$ and $T_{2}$ is an $\AA$-tree
      and the mappings $f_{1}:V(T_{1})\to V(T_{1,2})$ and
      $f_{2}:V(T_{2})\to V(T_{1,2})$ are weak topological embeddings.
    \end{itemize}
    \end{theorem}

    \begin{proof}
	
      (i)$\Longrightarrow$(ii) Assume that $T_{1}$ and $T_{2}$ are
      ancestrally compatible, and let $f_{1}:T_{1}\to T$ and
      $f_{2}:T_{2}\to T$ be two weak topological embeddings.  To prove
      that they are locally compatible, we shall show that they satisfy
      conditions (C1) and (C2).
      
      (C1) Assume that $T_{1}$ contains a path $v_{A}\cami v_{B}$. Since
      $f_{1}$ preserves this path, there exists a path $v_{A}\cami v_{B}$
      in $T$, and then this path must be reflected by $f_{2}$, yielding a
      path $v_{A}\cami v_{B}$ in $T_{2}$.
      
      (C2) Let $A,B,C\in \AA(T_{1})=\AA(T_{2})$.  Let
      $$
      y=v_{T_{1},A,B}\quad\mbox{ and }\quad z=v_{T_{1},B,C},
      $$
      and assume that there is a non-trivial path $z\cami y$; see
      Fig.~\ref{fig:th1-1}.  In particular, $y$ cannot be an ancestor of
      $v_{C}$: otherwise, it would be a common ancestor of $v_{B}$ and
      $v_{C}$, which would entail a path from $y$ to $z$ that cannot
      exist.
      
      Moreover,
      $$
      z=v_{T_{1},A,C}.
      $$
      Indeed, there are paths $z\cami v_{A}$, through $y$, and $z\cami
      v_{C}$, and therefore $z$ is a common ancestor of $v_{A}$ and
      $v_{C}$.  Then, $v_{T_{1},A,C}$ must be a node in the path $z\cami
      v_{A}$.  Assume that it is an intermediate node of this path.  If it
      is an intermediate node of the path $z\cami y$, then it will be a
      common ancestor of $v_{B}$, through $y$, and $v_{C}$, and therefore
      $z$ cannot be the most recent common ancestor of these two nodes.
      And if $v_{T_{1},A,C}$ is a node of the path $y\cami v_{A}$, then
      $y$ will be an ancestor of $v_{C}$, something that, as we have seen
      above, cannot happen.

    \begin{figure}[htb]
\begin{center}
        \begin{picture}(120,60)(-40,-10)
        \put(0,0){\line(1,1){40}} 
        \put(40,0){\line(-1,1){20}} 
        \put(40,40){\line(1,-1){40}} 
        \put(0,0){\circle*{3}} 
        \put(40,0){\circle*{3}} 
        \put(80,0){\circle*{3}} 
        \put(20,20){\circle*{3}} 
        \put(40,40){\circle*{3}} 
        \put(45,40){\makebox(0,0)[l]{$z=v_{B,C}=v_{A,C}$}}
        \put(0,-5){\makebox(0,0)[t]{$A$}}
        \put(40,-5){\makebox(0,0)[t]{$B$}}
        \put(80,-5){\makebox(0,0)[t]{$C$}}
        \put(15,20){\makebox(0,0)[r]{$y=v_{A,B}$}}
        \end{picture}
\end{center}
    \caption{\label{fig:th1-1}%
      The structure of $T_{1}$ above $v_{A},v_{B},v_{C}$.  The edges
      represent paths; any one of them can be trivial, except the path
      $z\cami y$, which is non-trivial by assumption}
    \end{figure}
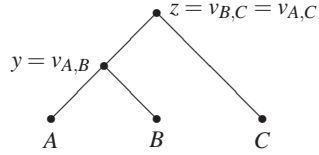

    Let us move now to $T$. Since $f_{1}$ preserves paths, $f_{1}(y)$ is a
    common ancestor of $v_{A}$ and $v_{B}$ and $f_{1}(z)$ is a common
    ancestor of $v_{B}$ and $v_{C}$, and there is a non-trivial path from
    $f_{1}(z)$ to $f_{1}(y)$.  Let
    $$
    y'=v_{T,A,B}\quad\mbox{ and }\quad z'=v_{T,B,C}.
    $$
    Then, $T$ contains paths $f_{1}(y)\cami y'$ and $f_{1}(z)\cami z'$,
    and it turns out that there is a non-trivial path $z'\cami f_{1}(y)$.
    Indeed, there are paths from $z'$ and from $f_{1}(y)$ to $v_{B}$, and
    therefore there must exist either a non-trivial path $z'\cami
    f_{1}(y)$ or a path $f_{1}(y)\cami z'$; but the latter cannot exist,
    because if it existed, then composing it with $z'\cami v_{C}$ we would
    obtain a path $f_{1}(y)\cami v_{C}$ that, when reflected by $f_{1}$,
    would entail a path $y\cami v_{C}$ in $T_{1}$ that does not exist.

    In particular, there is a non-trivial path $z'\cami y'$ in $T$.
    Arguing as in $T_{1}$, this implies that $z'$ is also the most recent
    common ancestor of $v_{A}$ and $v_{C}$ in $T$.  See Fig.~\ref{fig:T}
    for a representation of the structure of $T$ between $f_{1}(z)$ and
    $v_{A},v_{B},v_{C}$.

    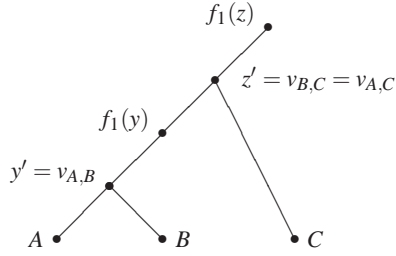
\begin{figure}[htb]
         \begin{center}
	  \begin{picture}(145,90)(-35,-5)
	    \put(0,0){\circle*{3}} 
	     \put(40,0){\circle*{3}} 
	     \put(20,20){\circle*{3}} 
	     \put(40,40){\circle*{3}} 
	     \put(60,60){\circle*{3}} 
	     \put(20,20){\circle*{3}} 
	     \put(90,0){\circle*{3}} 
	     \put(80,80){\circle*{3}} 
	  \put(0,0){\line(1,1){20}} 
	  \put(40,0){\line(-1,1){20}} 
	  \put(20,20){\line(1,1){20}} 
	  \put(40,40){\line(1,1){20}} 
	  \put(60,60){\line(1,1){20}} 
	  \put(60,60){\line(1,-2){30}} 
	   \put(-5,0){\makebox(0,0)[r]{$A$}}
	  \put(45,0){\makebox(0,0)[l]{$B$}}
	  \put(95,0){\makebox(0,0)[l]{$C$}}
	  \put(15,25){\makebox(0,0)[r]{$y'=v_{A,B}$}}
	  \put(35,45){\makebox(0,0)[r]{$f_{1}(y)$}}
	  \put(70,60){\makebox(0,0)[l]{$z'=v_{B,C}=v_{A,C}$}}
	  \put(75,85){\makebox(0,0)[r]{$f_{1}(z)$}}
	  \end{picture}
     \end{center}
    \caption{\label{fig:T}%
      The structure of $T$ above $v_{A},v_{B},v_{C}$.  The edges represent
      paths; any one of them can be trivial, except the path $z'\cami
      f_{1}(y)$, which is non-trivial}
    \end{figure}

    Consider finally the $\AA$-tree $T_{2}$, and set $x=v_{T_{2},B,C}$.
    Then, $f_{2}(x)$ will be a common ancestor of $v_{B}$ and $v_{C}$ in
    $T$ and therefore there will be a path $f_{2}(x)\cami z'$.  Composing
    this path with $z'\cami v_{A}$ we obtain a path $f_{2}(x)\cami v_{A}$
    which entails, since $f_{2}$ reflects paths, the existence of a path
    $x\cami v_{A}$.  Therefore, $x$ is also an ancestor of $v_{A}$, and
    thus there exists a path $x\cami v_{T_{2},A,B}$.  But then, there
    cannot exist a non-trivial path $v_{T_{2},A,B}\cami x$.

    This finishes the proof that $T_{1}$ and $T_{2}$ satisfy condition
    (C2).  \medskip

    (ii)$\Longrightarrow$(iii) Assume that $T_{1}$ and $T_{2}$ satisfy
    conditions (C1) and (C2).

    Let $A\in \AA(T_{1})=\AA(T_{2})$.  The smallest members of
    $\CC_{\AA}(T_{1})$ and $\CC_{\AA}(T_{2})$ containing $A$ are, of
    course, $\AA(v_{T_{1},A})$ and $\AA(v_{T_{2},A})$, respectively.  Now,
    the inequality $\AA(v_{T_{1},A})\neq \AA(v_{T_{2},A})$ violates
    property (C1): if, say, there exists a label $B\in \AA(v_{T_{1},A})-
    \AA(v_{T_{2},A})$, then $T_{1}$ contains a path $v_{A}\cami v_{B}$ but
    $T_{2}$ does not contain the corresponding path $v_{A}\cami v_{B}$.
    This proves the first condition in point (iii).

    Let now $X=\AA_{T_{1}}(x)\in \CC_{\AA}(T_{1})$ and
    $Y=\AA_{T_{2}}(y)\in \CC_{\AA}(T_{2})$ be such that $X\cap Y\neq
    \emptyset$, say $B\in X\cap Y$.  If none of them is included into the
    other one, then there exist labels $A\in X- Y$ and $C\in Y- X$.  Then,
    $C\notin \AA(v_{T_{1},A,B})$, because, since $x$ is a common ancestor
    of $v_{A}$ and $v_{B}$, there is a path $x\cami v_{T_{1},A,B}$ that
    entails the inclusion $\AA(v_{T_{1},A,B})\subseteq \AA(x)$, and by
    assumption $C\notin\AA(x)$.  Therefore, $v_{T_{1},B,C}$ is ``above''
    $v_{T_{1},A,B}$, that is, there exists a non-trivial path from
    $v_{B,C}$ to $v_{T_{1},A,B}$: since $B\in \AA(v_{T_{1},A,B})\cap
    \AA(v_{T_{1},B,C}$, if this path does not exist, then there must exist
    a path $v_{T_{1},A,B}\cami v_{T_{1},B,C}$ that will entail that $C\in
    \AA(v_{T_{1},A,B})$.

    In a similar way, we have that $A\notin \AA(v_{T_{2},B,C})$ and this
    entails a path $v_{T_{2},A,B}\cami v_{T_{2},B,C}$ in $T_{2}$.

    In all, if there exist $X\in \CC_{\AA}(T_{1})$ and $Y\in
    \CC_{\AA}(T_{2})$ such that $X\cap Y\neq \emptyset$, but
    $X\not\subseteq Y$ and $Y\not\subseteq X$, then there exist three
    labels $A,B,C\in \AA(T_{1})\cap \AA(T_{2})$ and non-trivial paths
    $v_{T_{1},B,C}\cami v_{T_{1},A,B}$ in $T_{1}$ and $v_{T_{2},A,B}\cami
    v_{T_{2},B,C}$ in $T_{2}$, which would contradict the assumption that
    $T_{1}$ and $T_{2}$ satisfy condition (C2).  \medskip

    (iii)$\Longrightarrow$(iv) Assume that $T_{1}$ and $T_{2}$ satisfy the
    conditions stated in point (iii).  Notice that the first condition in
    (iii) entails that $\LL(T_{1})=\LL(T_{2})$, because labels of leaves
    in an $\AA$-tree are characterized by the fact that the smallest
    member of the cluster representation containing the label is a
    singleton.

    To simplify the notations, we shall denote the join of $T_{1}$ and
    $T_{2}$ by simply $T$. In this case, since $\AA(T_{1})=\AA(T_{2})$,
    this join $T$ is obtained using the construction given in
    Definition~\ref{T12}.(a).  Let us check that it is an $\AA$-tree:
    \begin{itemize}
    \item [$\bullet$] It is clear that its leaves are the nodes of the
      form $w_{A,1}$, and they are labeled.
    \item The nodes of $T$ are injectively labeled: it is impossible the
      existence of two different sets of labels $Y_{1},Y_{2}\in \CC$ such
      that
      $$
      Y_{1}=(\bigcup \{Z\in \CC\mid Z\subsetneq
      Y_{1}\})\sqcup\{A\},\quad Y_{2}=(\bigcup \{Z\in \CC\mid Z\subsetneq
      Y_{2}\})\sqcup\{A\},
      $$
      because in this case $Y_{1}\cap Y_{2}\neq \emptyset$ and
      therefore $Y_{1}\subsetneq Y_{2}$ or $Y_{2}\subsetneq Y_{1}$, which
      would entail that one of them contains a member of $\CC$ that
      already contains $A$.
      
      As we shall see below, $\AA(T)=\AA(T_{1})=\AA(T_{2})$.
      
    \item[$\bullet$] It is a tree.  To prove it, assume first that a node
      $w_{Z,j}$ has two parents.  Then, by construction, it must happen
      that $j=n_{Z}$ and then the parents are nodes $w_{Y_{1},1}$ and
      $w_{Y_{2},1}$ with $Y_{1},Y_{2}\in \CC$, $Y_{1}\neq Y_{2}$, such
      that $Z\subsetneq Y_{1}$, $Z\subsetneq Y_{2}$ and in both cases such
      that no other member of $\CC$ lies strictly between $Z$ and the
      corresponding $Y_{i}$.  But then $Y_{1}\cap Y_{2}\neq \emptyset$ and
      therefore $Y_{1}\subseteq Y_{2}$ or $Y_{2}\subseteq Y_{1}$: if
      $Y_{1},Y_{2}\in \CC_{\AA}(T_{1})$ or $Y_{1},Y_{2}\in
      \CC_{\AA}(T_{2})$, by Lemma~\ref{lem-int1}, and if each one of them
      belongs to a different cluster representation, by assumption.  This
      forbids that both $Y_{1}$ and $Y_{2}$ are minimal over $Z$.
      Therefore, each $w_{Z,j}$ can have only one parent.
      
      Now, if $X,Y\in \CC$ and $Y\subseteq X$, there is a unique path
      $w_{X,i}\cami w_{Y,j}$ for every $i=1,\ldots,n_{X}$ and
      $j=1,\ldots,n_{Y}$ (if $X=Y$, then this happens for every $1\leq
      j\leq i\leq n_{X}$). If $X=Y$, it is obvious by construction, and
      when $Y\subsetneq X$, if
      $$
      Y\subsetneq Z_{1}\subsetneq Z_{2}\subsetneq \cdots \subsetneq
      Z_{k}\subsetneq X
      $$
      is a maximal chain of sets of labels between $Y$ and $X$ with
      $Z_{1},\ldots,Z_{k}\in \CC$, then this path is obtained as the
      composition of paths
      $$
      w_{X,i}\cami w_{X,1}\cami w_{Z_{k},n_{Z_{k}}}\cami
      w_{Z_{k},1}\cami w_{Z_{k-1},n_{Z_{k-1}}}\cami\cdots \cami
      w_{Z_{1},1}\cami w_{Y,n_{Y}}\cami w_{Y,j}.
      $$
      And this path is unique because every node has at most one
      parent.
      
      Then, since $\AA(T_{1})=\AA(T_{2})\in \CC$, because it is the
      cluster of the roots of both trees, every node $w_{Y,j}$ is a
      descendant of $w_{\AA(T_{1}),1}$, that is, $w_{\AA(T_{1}),1}$ is the
      root of $T$.
    \end{itemize}

    This $\AA$-tree $T$ satisfies the following properties that we shall
    use below:
    \begin{itemize}
       
    \item [$\bullet$] $\AA(w_{Y,j})=Y$, for every node $w_{Y,j}$.
      
      This is easily proved by algebraic induction over the structure of
      $T$.  If $Y=\{A\}$ and $j=1$, then $w_{Y,1}$ is a leaf of $T$
      labeled $A$, while if $Y=\{A\}$ and $j>1$, then the only labeled
      descendant of $w_{Y,j}$ in $T$ is the leaf $w_{Y,1}$.  Thus,
      $\AA(w_{A,j})=\{A\}$ for every $A\in \LL(A_{1})=\LL(A_{2})$ and
      $j=1,\ldots,n_{A}$.
      
      Now assume that $\AA(w_{Z,j})=Z$ for every $Z\subsetneq Y$ and
      $j=1,\ldots,n_{Z}$, and let us prove it for $Y$ and every
      $j=1,\ldots,n_{Y}$.  If $j=1$, then the children of $w_{Y,1}$ are
      the nodes $w_{Z,n_{Z}}$ with $Z\subsetneq Y$ and maximal with this
      property. And then, if $w_{Y,1}$ is not labeled,
      $$
    \begin{array}{rl}
    \AA(w_{Y,1}) & =\bigcup\{\AA(w_{Z,n_{Z}})\mid Z\subsetneq  Y\mbox{ and
    maximal with this property}\}\\
    & =\bigcup\{\AA(w_{Z,n_{Z}})\mid Z\subsetneq  Y\} =\bigcup\{Z\mid Z\subsetneq  Y\}=Y
    \end{array}
    $$
    (in the second equality we use that if $Z\subsetneq Y$, then there
    exists some maximal $Z_{0}\subsetneq Y$ such that $Z\subseteq Z_{0}$,
    and then there exists a path $w_{Z_{0},1}\cami w_{Z,1}$ that entails
    that $\AA(w_{Z,1})\subseteq \AA(w_{Z_{0},1})$), while, if $w_{Y,1}$ is
    labeled, say with label $A$, then
    $$
    \begin{array}{rl}
    \AA(w_{Y,1}) & \!\! =(\bigcup\{\AA(w_{Z,n_{Z}})\mid Z\subsetneq  Y\mbox{ and
    maximal with this property}\})\sqcup\{A\}\\
    & \!\! =(\{\AA(w_{Z,n_{Z}})\mid Z\subsetneq  Y\})\sqcup\{A\} 
    =(\bigcup\{Z\mid Z\subsetneq  Y\})\sqcup\{A\}=Y.
    \end{array}
    $$

    Finally, if $j>1$, then there is a path $w_{Y,j}\cami w_{Y,1}$ with
    the origin and all its intermediate nodes elementary and unlabeled,
    and therefore $\AA(w_{Y,j})=\AA(w_{Y,1})=Y$.

    \item [$\bullet$] In particular, $w_{Y,1}=v_{T,Y}$, for every $Y\in
      \CC$, because, as we have just proved, $\AA(w_{Y,1})=Y$, and all
      children $w_{Z,n_{Z}}$ of $w_{Y,1}$ are such that
      $\AA(w_{Z,n_{Z}})=Z\subsetneq Y$.
    \end{itemize}

    Let us prove now that $f_{1}:V(T_{1})\to V(T)$ is a weak topological
    embedding $f_{1}:T_{1}\to T$; by symmetry, it will be true also for
    $T_{2}$.

    Let us check that $f_{1}$ preserves labels.  Let $A\in\AA(T_{1})$ and
    $Y=\AA(v_{T_{1},A})$.  Then, in particular, and using the notations of
    Definition~\ref{T12}, $v_{T_{1},A}=v_{T_{1},Y}=x^{(1)}_{Y,1}$, and
    hence $f_{1}(v_{T_{1},A})=w_{Y,1}$.  We must check that this node has
    label $A$, that is, that
    $$
    Y=(\bigcup \{Z\in \CC\mid Z\subsetneq Y\})\sqcup\{A\},
    $$
    because in this case, and only in this case, $w_{Y,1}$ is labeled
    $A$.  

    So, assume that there exists some $Z\in \CC$ such that $Z\subsetneq Y$
    and $A\in Z$.  Such a $Z$ cannot belong to $\CC_{\AA}(T_{1})$, and
    therefore there exists some $z\in V(T_{2})$ such that $\AA(z)=Z$.
    Since $A\in \AA(z)$, there exists a path $z\cami v_{T_{2},A}$ in
    $T_{2}$ and therefore $\AA(v_{T_{2},A})\subseteq \AA(z)$.  But, by the
    first condition in (iii), $\AA(v_{A})=Y$ and therefore this inequality
    says $Y\subseteq Z$, which is impossible.  Therefore, $A\notin Z$ for
    every $Z\subsetneq Y$, as we wanted to have.

    Finally, let us prove that $f_{1}$ preserves and reflects paths.  Let
    $u \cami v$ be a non-trivial path in $T_{1}$, so that $\AA(v)\subseteq
    \AA(u)$.  If $\AA(v)= \AA(u)$, then $u=x^{(1)}_{\AA(v),i}$ and
    $v=x^{(1)}_{\AA(v),j}$ with $i>j$, and then by construction $T$
    contains a path from $f_{1}(u)=w_{\AA(v),i}$ to
    $f_{1}(v)=w_{\AA(v),j}$.  If, on the contrary, $\AA(v)\subsetneq
    \AA(u)$, then $f_{1}(u)=w_{\AA(u),i}$ and $f_{1}(v)=w_{\AA(v),j}$ for
    some $i,j$, and, as we saw when we proved that $T$ is an $\AA$-tree,
    $T$ contains a path $w_{\AA(u),i}\cami w_{\AA(v),j}$.

    Conversely, let $f_{1}(u) \cami f_{1}(v)$ be a path in $T$, and assume
    that $f_{1}(u)=w_{\AA(u),i}$ and $f_{1}(v)=w_{\AA(v),j}$.  Then, the
    existence of this path entails that
    $$
    \AA(v)=\AA(w_{\AA(v),j})\subseteq \AA(w_{\AA(u),i})=\AA(u).
    $$
    If this inclusion is strict, then Corollary~\ref{lem-int2} implies
    the existence of a path $u \cami v$ in $T_{1}$.  On the other hand, if
    $\AA(v)= \AA(u)$, then $u=x^{(1)}_{\AA(u),i}$ and
    $v=x^{(1)}_{\AA(u),j}$ for some $1\leq i,j\leq m_{1,\AA(u)}$, and then
    the definition of $f_{1}$ implies that if $T$ contains a path
    $f_{1}(u) \cami f_{1}(v)$, then $i>j$ and therefore there is a path
    $u\cami v$ in $T_{1}$.

    This finishes the proof that $f_{1}:T_{1}\to T$ is a weak topological
    embedding.

    (iv)$\Longrightarrow$(i) This implication is obvious.  \qed
    \end{proof}

    \begin{corollary}
      Let $T_{1}$ and $T_{2}$ be $\AA$-trees.  Then, the following
      assertions are equivalent:
    \begin{itemize}
    \item[(i)] $T_{1}$ and $T_{2}$ are ancestrally compatible.
    \item[(ii)] $T_{1}$ and $T_{2}$ are locally compatible.
    \item[(iii)] Their $\AA$-subtrees $\bar{T}_{1}$ and $\bar{T}_{2}$
      described in Definition~\ref{def:hat} satisfy condition (iii) in
      Theorem~\ref{th:ancestrally comp}.
    \item[(iv)] The join $T_{1,2}$ of $T_{1}$ and $T_{2}$ is an $\AA$-tree
      and the mappings $f_{1}:V(T_{1})\to V(T_{1,2})$ and
      $f_{2}:V(T_{2})\to V(T_{1,2})$ are weak topological embeddings.
    \end{itemize}
    \end{corollary}

    \begin{proof}
      By Lemma~\ref{comp-hat}, $T_{1}$ and $T_{2}$ are locally compatible
      if and only if $\bar{T}_{1}$ and $\bar{T}_{2}$ are so, and by
      Proposition~\ref{lem-restr}, $T_{1}$ and $T_{2}$ are ancestrally
      compatible if and only if $\bar{T}_{1}$ and $\bar{T}_{2}$ are so.
      These facts, together with the last theorem, prove the implications
      (i)$\Rightarrow$(ii) and (ii)$\Rightarrow$(iii). As far as
      (iii)$\Rightarrow$(iv) goes, it is a direct consequence of the
      corresponding implication in the last theorem together with the
      proof of Proposition~\ref{lem-restr}.  \qed
    \end{proof}

    \begin{corollary}
      Let $T_{1}$ and $T_{2}$ be semi-labeled trees over $\AA$.  Then, the
      following assertions are equivalent:
    \begin{itemize}
    \item[(i)] $T_{1}$ and $T_{2}$ admit simultaneous weak topological
      embeddings into a same semi-labeled tree over $\AA$.
    \item[(ii)] $T_{1}$ and $T_{2}$ are ancestrally compatible.
    \item[(iii)] $T_{1}$ and $T_{2}$ are locally compatible.
    \item[(iv)] Their $\AA$-subtrees $\bar{T}_{1}$ and $\bar{T}_{2}$
      described in Definition~\ref{def:hat} satisfy condition (iii) in
      Theorem~\ref{th:ancestrally comp}.
    \item[(v)] The join $T_{1,2}$ of $T_{1}$ and $T_{2}$ is a semi-labeled
      tree and the mappings $f_{1}:V(T_{1})\to V(T_{1,2})$ and
      $f_{2}:V(T_{2})\to V(T_{1,2})$ are weak topological embeddings.
    \end{itemize}
    \end{corollary}

    \begin{proof}
      It only remains to prove (iv)$\Longrightarrow$(v). And to do that,
      it is enough to notice that if $T_{1}$ and $T_{2}$ are semi-labeled
      trees over $\AA$ such that $\bar{T}_{1}$ and $\bar{T}_{2}$ satisfy
      condition (iii) in Theorem~\ref{th:ancestrally comp}, then their
      join $T_{1,2}$ is not only an $\AA$-tree, but a semi-labeled tree,
      because, since $f_{1}:T_{1}\to T_{1,2}$ and $f_{2}:T_{2}\to T_{1,2}$
      are jointly surjective, no elementary node in it remains unlabeled.
      \qed
    \end{proof}

    \section{Algorithmic Details}
    \label{sect:implement}

    The equivalence between ancestral compatibility and the properties of
    the cluster representations of the trees established in
    Theorem~\ref{th:ancestrally comp}, leads to a very simple
    polynomial-time algorithm for testing ancestral compatibility of two
    semi-labeled trees. The detailed pseudo-code of the algorithm is shown
    in Fig.~\ref{fig:compat}.

    \begin{figure}[htb]
\begin{center}
         \begin{algorithm2e}[H]
        \SetFuncSty{emph}
        \SetArgSty{textrm}
        \dontprintsemicolon
        compatible($T_1,T_2$)\;
        $\AA := \AA(T_1) \cap \AA(T_2)$\;
        $\bar{T}_1 := T_1|\AA$\;
        $\bar{T}_2 := T_2|\AA$\;
        \ForEach{label $A \in \AA$}{
          let  $X_1$ be the smallest member of $\CC_\AA(\bar{T}_1)$ containing $A$\;
          let  $X_2$ be the smallest member of $\CC_\AA(\bar{T}_2)$ containing $A$\;
          \If{$X_1 \neq X_2$}{
        \Return{$X_1$ and $X_2$ are incompatible}
          }
        }
        \ForEach{cluster $X_1 \in \CC_\AA(\bar{T}_1)$}{
          \ForEach{cluster $X_2 \in \CC_\AA(\bar{T}_2)$}{
        \If{$X_1 \cap X_2 \neq \emptyset$ and $X_1 \not\subseteq X_2$ and $X_2 \not\subseteq X_1$}{
          \Return{$X_1$ and $X_2$ are incompatible}
        }
          }
        }
        \Return{$T_1$ and $T_2$ are compatible}
        \end{algorithm2e}
\end{center}
     \caption{\label{fig:compat}%
      Algorithm for testing ancestral compatibility of two semi-labeled
      trees $T_1$ and $T_2$}
    \end{figure}

    We have implemented in Perl this compatibility test, and the
    implementation is freely available for download from the BioPerl
    collection of Perl modules for computational
    biology~\cite{stajich.ea:2002}. Given two semi-labeled trees $T_1$ and
    $T_2$ with common labels $\AA = \AA(T_1) \cap \AA(T_2)$, if the trees
    are incompatible, the actual implementation collects and returns all
    labels $A \in \AA$ such that be the smallest member of
    $\CC_\AA(T_1|\AA)$ containing $A$ does not coincide with be the
    smallest member of $\CC_\AA(T_2|\AA)$ containing $A$, as well as all
    pairs of clusters $X_1 \in \CC_\AA(T_1|\AA)$ and $X_2 \in
    \CC_\AA(T_2|\AA)$ such that $X_1 \cap X_2 \neq \emptyset$, $X_1
    \not\subseteq X_2$, and $X_2 \not\subseteq X_1$. This additional
    information constitutes a \emph{certificate of incompatibility}, which
    can be useful for checking the underlying phylogenetic studies that
    have lead to incompatible clusters.

    The following Perl code illustrates the use of the
    \texttt{Bio::Tree::Compatible} module for testing compatibility of two
    semi-labeled trees and listing all pairs of incompatible clusters in
    the trees.

    \begin{lstlisting}
    use Bio::Tree::Compatible;
    use Bio::TreeIO;

    my $filename = $ARGV[0];
    my $input = new Bio::TreeIO('-format' => 'newick',
				'-file' => $filename);
    my $t1 = $input->next_tree;
    my $t2 = $input->next_tree;

    my ($incompat, $ilabels, $inodes) =
      $t1->Bio::Tree::Compatible::is_compatible($t2);

    if ($incompat) {
      print "the trees are incompatible\n";

      my %cluster1 = %{
	$t1->Bio::Tree::Compatible::cluster_representation };
      my %cluster2 = %{
	$t2->Bio::Tree::Compatible::cluster_representation };

      if (scalar(@$ilabels)) {
	foreach my $label (@$ilabels) {
	  my $node1 = $t1->find_node(-id => $label);
	  my $node2 = $t2->find_node(-id => $label);
	  my @c1 = sort @{ $cluster1{$node1} };
	  my @c2 = sort @{ $cluster2{$node2} };
	  print "label $label";
	  print " cluster"; map { print " ",$_ } @c1;
	  print " cluster"; map { print " ",$_ } @c2;
	  print "\n";
	}
      }

      if (scalar(@$inodes)) {
	while (@$inodes) {
	  my $node1 = shift @$inodes;
	  my $node2 = shift @$inodes;
	  my @c1 = sort @{ $cluster1{$node1} };
	  my @c2 = sort @{ $cluster2{$node2} };
	  print "cluster"; map { print " ",$_ } @c1;
	  print " properly intersects cluster";
	  map { print " ",$_ } @c2; print "\n";
	}
      }
    } else {
      print "the trees are compatible\n";
    }
    \end{lstlisting}

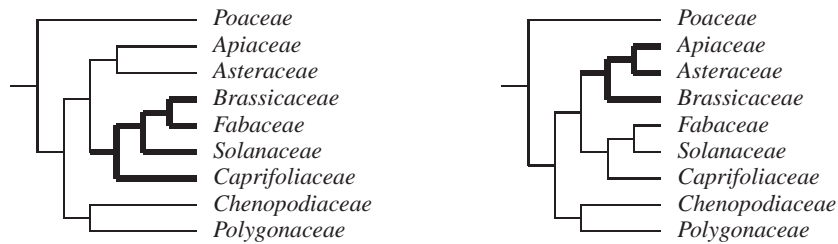
\begin{figure}[htb]
    \begin{center}
    \setlength{\unitlength}{.5pt}
    \small
    \itshape
    \begin{picture}(260,180)(0,0)
    \put(020,170){\line(1,0){120}}
    \put(080,150){\line(1,0){060}}
    \put(080,130){\line(1,0){060}}
    \linethickness{2pt} 
    \put(120,110){\line(1,0){020}}
    \put(120,090){\line(1,0){020}}
    \put(100,070){\line(1,0){040}}
    \put(080,050){\line(1,0){060}}
    \put(060,070){\line(1,0){020}}
    \put(080,090){\line(1,0){020}}
    \put(100,100){\line(1,0){020}}
    \put(080,050){\line(0,1){040}}
    \put(100,070){\line(0,1){030}}
    \put(120,090){\line(0,1){020}}
    \thinlines
    \put(060,030){\line(1,0){080}}
    \put(060,010){\line(1,0){080}}
    \put(000,120){\line(1,0){020}}
    \put(020,070){\line(1,0){020}}
    \put(040,110){\line(1,0){020}}
    \put(040,020){\line(1,0){020}}
    \put(060,140){\line(1,0){020}}
    \put(020,070){\line(0,1){100}}
    \put(040,020){\line(0,1){090}}
    \put(060,070){\line(0,1){070}}
    \put(060,010){\line(0,1){020}}
    \put(080,130){\line(0,1){020}}
    \put(145,160){\makebox(60,20)[l]{\phantom{y}Poaceae}}
    \put(145,140){\makebox(60,20)[l]{\phantom{y}Apiaceae}}
    \put(145,120){\makebox(60,20)[l]{\phantom{y}Asteraceae}}
    \put(145,100){\makebox(60,20)[l]{\phantom{y}Brassicaceae}}
    \put(145,080){\makebox(60,20)[l]{\phantom{y}Fabaceae}}
    \put(145,060){\makebox(60,20)[l]{\phantom{y}Solanaceae}}
    \put(145,040){\makebox(60,20)[l]{\phantom{y}Caprifoliaceae}}
    \put(145,020){\makebox(60,20)[l]{\phantom{y}Chenopodiaceae}}
    \put(145,000){\makebox(60,20)[l]{\phantom{y}Polygonaceae}}
    \end{picture}
    \hfil
    \begin{picture}(240,180)(0,0)
    \put(020,170){\line(1,0){100}}
    \linethickness{2pt} 
    \put(100,150){\line(1,0){020}}
    \put(100,130){\line(1,0){020}}
    \put(080,110){\line(1,0){040}}
    \put(060,130){\line(1,0){020}}
    \put(080,140){\line(1,0){020}}
    \put(080,110){\line(0,1){030}}
    \put(100,130){\line(0,1){020}}
    \thinlines
    \put(100,090){\line(1,0){020}}
    \put(100,070){\line(1,0){020}}
    \put(080,050){\line(1,0){040}}
    \put(060,030){\line(1,0){060}}
    \put(060,010){\line(1,0){060}}
    \put(000,120){\line(1,0){020}}
    \put(020,060){\line(1,0){020}}
    \put(040,100){\line(1,0){020}}
    \put(040,020){\line(1,0){020}}
    \put(060,070){\line(1,0){020}}
    \put(080,080){\line(1,0){020}}
    \put(020,060){\line(0,1){110}}
    \put(040,020){\line(0,1){080}}
    \put(060,070){\line(0,1){060}}
    \put(060,010){\line(0,1){020}}
    \put(080,050){\line(0,1){030}}
    \put(100,070){\line(0,1){020}}
    \put(125,160){\makebox(60,20)[l]{\phantom{y}Poaceae}}
    \put(125,140){\makebox(60,20)[l]{\phantom{y}Apiaceae}}
    \put(125,120){\makebox(60,20)[l]{\phantom{y}Asteraceae}}
    \put(125,100){\makebox(60,20)[l]{\phantom{y}Brassicaceae}}
    \put(125,080){\makebox(60,20)[l]{\phantom{y}Fabaceae}}
    \put(125,060){\makebox(60,20)[l]{\phantom{y}Solanaceae}}
    \put(125,040){\makebox(60,20)[l]{\phantom{y}Caprifoliaceae}}
    \put(125,020){\makebox(60,20)[l]{\phantom{y}Chenopodiaceae}}
    \put(125,000){\makebox(60,20)[l]{\phantom{y}Polygonaceae}}
    \end{picture}
    \end{center}
    \caption{\label{fig:angiosperm}%
      Two incompatible phylogenetic trees, obtained from study
      S11x5x95c19c35c30 in TreeBASE. The clusters shown with thick lines
      are incompatible.}
    \end{figure}

    An application of \texttt{Bio::Tree::Compatible} is shown in
    Fig.~\ref{fig:angiosperm}. The input consists of two phylogenetic
    trees describing the evolution of angiosperms (plants that flower and
    form fruits with seeds), obtained from study S11x5x95c19c35c30 in the
    TreeBASE~\cite{morell:1996} phylogenetic database.

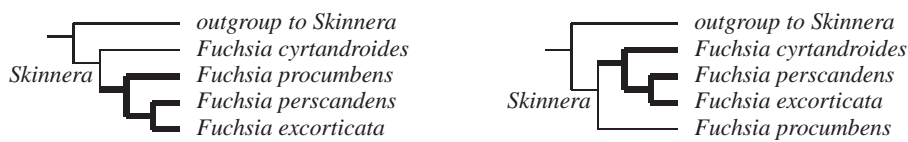
\begin{figure}[htb]
    \begin{center}
    \setlength{\unitlength}{.5pt}
    \small
    \itshape
    \begin{picture}(260,100)(0,0)
    \linethickness{2pt} 
    \put(040,040){\line(1,0){020}}
    \put(060,050){\line(1,0){040}}
    \put(060,020){\line(1,0){020}}
    \put(080,030){\line(1,0){020}}
    \put(080,010){\line(1,0){020}}
    \put(060,020){\line(0,1){030}}
    \put(080,010){\line(0,1){020}}
    \thinlines
    \put(000,080){\line(1,0){020}}
    \put(020,090){\line(1,0){080}}
    \put(020,060){\line(1,0){020}}
    \put(040,070){\line(1,0){060}}
    \put(020,060){\line(0,1){030}}
    \put(040,040){\line(0,1){030}}
    \put(105,080){\makebox(60,20)[l]{\phantom{y}outgroup to Skinnera}}
    \put(105,060){\makebox(60,20)[l]{\phantom{y}Fuchsia cyrtandroides}}
    \put(105,040){\makebox(60,20)[l]{\phantom{y}Fuchsia procumbens}}
    \put(105,020){\makebox(60,20)[l]{\phantom{y}Fuchsia perscandens}}
    \put(105,000){\makebox(60,20)[l]{\phantom{y}Fuchsia excorticata}}
    \put(-25,040){\makebox(60,20)[r]{\phantom{y}Skinnera}}
    \end{picture}
    \quad\hfil
    \begin{picture}(260,100)(0,0)
    \linethickness{2pt} 
    \put(040,060){\line(1,0){020}}
    \put(060,040){\line(1,0){020}}
    \put(060,070){\line(1,0){040}}
    \put(080,050){\line(1,0){020}}
    \put(080,030){\line(1,0){020}}
    \put(060,040){\line(0,1){030}}
    \put(080,030){\line(0,1){020}}
    \thinlines
    \put(000,070){\line(1,0){020}}
    \put(020,090){\line(1,0){080}}
    \put(020,040){\line(1,0){020}}
    \put(040,010){\line(1,0){060}}
    \put(020,040){\line(0,1){050}}
    \put(040,010){\line(0,1){050}}
    \put(105,080){\makebox(60,20)[l]{\phantom{y}outgroup to Skinnera}}
    \put(105,060){\makebox(60,20)[l]{\phantom{y}Fuchsia cyrtandroides}}
    \put(105,040){\makebox(60,20)[l]{\phantom{y}Fuchsia perscandens}}
    \put(105,020){\makebox(60,20)[l]{\phantom{y}Fuchsia excorticata}}
    \put(105,000){\makebox(60,20)[l]{\phantom{y}Fuchsia procumbens}}
    \put(-25,020){\makebox(60,20)[r]{\phantom{y}Skinnera}}
    \end{picture}
    \end{center}
    \caption{\label{fig:skinnera}%
      Two incompatible semi-labeled trees, obtained from study
      S11x4x95c21c16c44 in TreeBASE. The clusters shown with thick lines
      are incompatible.}
    \end{figure}

    Another application of \texttt{Bio::Tree::Compatible} is shown in
    Fig.~\ref{fig:skinnera}. The input consists of two semi-labeled trees
    describing the evolution of Skinnera (a group of four Fuchsia species
    that grows spontaneously out of the American continent, in New Zealand
    and on Tahiti), obtained from study S11x4x95c21c16c44 in TreeBASE.

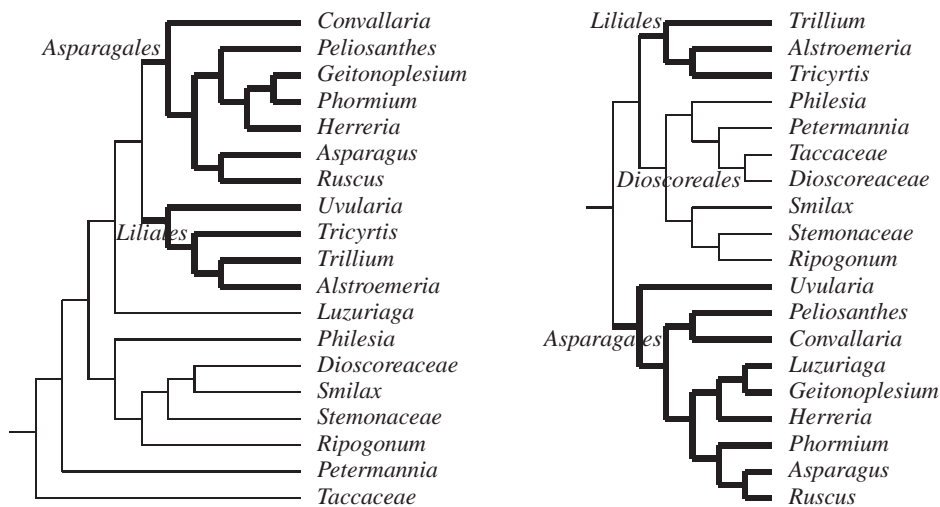
\begin{figure}[htb]
    \begin{center}
    \setlength{\unitlength}{.5pt}
    \small
    \itshape
    \begin{picture}(360,380)(0,0)
    \linethickness{2pt} 
    \put(200,330){\line(1,0){020}}
    \put(200,310){\line(1,0){020}}
    \put(180,320){\line(1,0){020}}
    \put(180,290){\line(1,0){040}}
    \put(160,350){\line(1,0){060}}
    \put(160,310){\line(1,0){020}}
    \put(160,270){\line(1,0){060}}
    \put(160,250){\line(1,0){060}}
    \put(160,190){\line(1,0){060}}
    \put(160,170){\line(1,0){060}}
    \put(140,330){\line(1,0){020}}
    \put(140,260){\line(1,0){020}}
    \put(140,210){\line(1,0){080}}
    \put(140,180){\line(1,0){020}}
    \put(120,370){\line(1,0){100}}
    \put(120,300){\line(1,0){020}}
    \put(120,230){\line(1,0){100}}
    \put(120,200){\line(1,0){020}}
    \put(100,340){\line(1,0){020}}
    \put(100,220){\line(1,0){020}}
    \put(120,200){\line(0,1){030}}
    \put(120,300){\line(0,1){070}}
    \put(140,180){\line(0,1){030}}
    \put(140,260){\line(0,1){070}}
    \put(160,170){\line(0,1){020}}
    \put(160,250){\line(0,1){020}}
    \put(160,310){\line(0,1){040}}
    \put(180,290){\line(0,1){030}}
    \put(200,310){\line(0,1){020}}
    \thinlines
    \put(140,110){\line(1,0){080}}
    \put(140,090){\line(1,0){080}}
    \put(120,100){\line(1,0){020}}
    \put(120,070){\line(1,0){100}}
    \put(100,090){\line(1,0){020}}
    \put(100,050){\line(1,0){120}}
    \put(080,280){\line(1,0){020}}
    \put(080,150){\line(1,0){140}}
    \put(080,130){\line(1,0){140}}
    \put(080,070){\line(1,0){020}}
    \put(060,220){\line(1,0){020}}
    \put(060,100){\line(1,0){020}}
    \put(040,160){\line(1,0){020}}
    \put(040,030){\line(1,0){180}}
    \put(020,100){\line(1,0){020}}
    \put(020,010){\line(1,0){200}}
    \put(000,060){\line(1,0){020}}
    \put(020,010){\line(0,1){090}}
    \put(040,030){\line(0,1){130}}
    \put(060,100){\line(0,1){120}}
    \put(080,070){\line(0,1){060}}
    \put(080,150){\line(0,1){130}}
    \put(100,050){\line(0,1){040}}
    \put(100,220){\line(0,1){120}}
    \put(120,070){\line(0,1){030}}
    \put(140,090){\line(0,1){020}}
    \put(225,360){\makebox(60,20)[l]{\phantom{y}Convallaria}}
    \put(225,340){\makebox(60,20)[l]{\phantom{y}Peliosanthes}}
    \put(225,320){\makebox(60,20)[l]{\phantom{y}Geitonoplesium}}
    \put(225,300){\makebox(60,20)[l]{\phantom{y}Phormium}}
    \put(225,280){\makebox(60,20)[l]{\phantom{y}Herreria}}
    \put(225,260){\makebox(60,20)[l]{\phantom{y}Asparagus}}
    \put(225,240){\makebox(60,20)[l]{\phantom{y}Ruscus}}
    \put(225,220){\makebox(60,20)[l]{\phantom{y}Uvularia}}
    \put(225,200){\makebox(60,20)[l]{\phantom{y}Tricyrtis}}
    \put(225,180){\makebox(60,20)[l]{\phantom{y}Trillium}}
    \put(225,160){\makebox(60,20)[l]{\phantom{y}Alstroemeria}}
    \put(225,140){\makebox(60,20)[l]{\phantom{y}Luzuriaga}}
    \put(225,120){\makebox(60,20)[l]{\phantom{y}Philesia}}
    \put(225,100){\makebox(60,20)[l]{\phantom{y}Dioscoreaceae}}
    \put(225,080){\makebox(60,20)[l]{\phantom{y}Smilax}}
    \put(225,060){\makebox(60,20)[l]{\phantom{y}Stemonaceae}}
    \put(225,040){\makebox(60,20)[l]{\phantom{y}Ripogonum}}
    \put(225,020){\makebox(60,20)[l]{\phantom{y}Petermannia}}
    \put(225,000){\makebox(60,20)[l]{\phantom{y}Taccaceae}}
    \put(055,340){\makebox(60,20)[r]{\phantom{y}Asparagales}}
    \put(075,200){\makebox(60,20)[r]{\phantom{y}Liliales}}
    \end{picture}
    \quad\hfil
    \begin{picture}(280,380)(0,0)
    \linethickness{2pt} 
    \put(080,350){\line(1,0){060}}
    \put(080,330){\line(1,0){060}}
    \put(060,370){\line(1,0){080}}
    \put(060,340){\line(1,0){020}}
    \put(040,360){\line(1,0){020}}
    \put(080,330){\line(0,1){020}}
    \put(060,340){\line(0,1){030}}
    \put(120,110){\line(1,0){020}}
    \put(120,090){\line(1,0){020}}
    \put(120,030){\line(1,0){020}}
    \put(120,010){\line(1,0){020}}
    \put(100,100){\line(1,0){020}}
    \put(100,070){\line(1,0){040}}
    \put(100,050){\line(1,0){040}}
    \put(100,020){\line(1,0){020}}
    \put(080,150){\line(1,0){060}}
    \put(080,130){\line(1,0){060}}
    \put(080,090){\line(1,0){020}}
    \put(080,040){\line(1,0){020}}
    \put(060,140){\line(1,0){020}}
    \put(060,070){\line(1,0){020}}
    \put(040,170){\line(1,0){100}}
    \put(040,110){\line(1,0){020}}
    \put(020,140){\line(1,0){020}}
    \put(120,090){\line(0,1){020}}
    \put(120,010){\line(0,1){020}}
    \put(100,070){\line(0,1){030}}
    \put(100,020){\line(0,1){030}}
    \put(080,130){\line(0,1){020}}
    \put(080,040){\line(0,1){050}}
    \put(060,070){\line(0,1){070}}
    \put(040,110){\line(0,1){060}}
    \thinlines
    \put(145,360){\makebox(60,20)[l]{\phantom{y}Trillium}}
    \put(145,340){\makebox(60,20)[l]{\phantom{y}Alstroemeria}}
    \put(145,320){\makebox(60,20)[l]{\phantom{y}Tricyrtis}}
    \put(145,300){\makebox(60,20)[l]{\phantom{y}Philesia}}
    \put(145,280){\makebox(60,20)[l]{\phantom{y}Petermannia}}
    \put(145,260){\makebox(60,20)[l]{\phantom{y}Taccaceae}}
    \put(145,240){\makebox(60,20)[l]{\phantom{y}Dioscoreaceae}}
    \put(145,220){\makebox(60,20)[l]{\phantom{y}Smilax}}
    \put(145,200){\makebox(60,20)[l]{\phantom{y}Stemonaceae}}
    \put(145,180){\makebox(60,20)[l]{\phantom{y}Ripogonum}}
    \put(145,160){\makebox(60,20)[l]{\phantom{y}Uvularia}}
    \put(145,140){\makebox(60,20)[l]{\phantom{y}Peliosanthes}}
    \put(145,120){\makebox(60,20)[l]{\phantom{y}Convallaria}}
    \put(145,100){\makebox(60,20)[l]{\phantom{y}Luzuriaga}}
    \put(145,080){\makebox(60,20)[l]{\phantom{y}Geitonoplesium}}
    \put(145,060){\makebox(60,20)[l]{\phantom{y}Herreria}}
    \put(145,040){\makebox(60,20)[l]{\phantom{y}Phormium}}
    \put(145,020){\makebox(60,20)[l]{\phantom{y}Asparagus}}
    \put(145,000){\makebox(60,20)[l]{\phantom{y}Ruscus}}
    \put(-05,360){\makebox(60,20)[l]{\phantom{y}Liliales}}
    \put(015,240){\makebox(60,20)[l]{\phantom{y}Dioscoreales}}
    \put(-40,120){\makebox(60,20)[l]{\phantom{y}Asparagales}}
    \put(120,270){\line(1,0){020}}
    \put(120,250){\line(1,0){020}}
    \put(100,290){\line(1,0){040}}
    \put(100,260){\line(1,0){020}}
    \put(100,210){\line(1,0){040}}
    \put(100,190){\line(1,0){040}}
    \put(080,310){\line(1,0){060}}
    \put(080,280){\line(1,0){020}}
    \put(080,230){\line(1,0){060}}
    \put(080,200){\line(1,0){020}}
    \put(060,300){\line(1,0){020}}
    \put(060,220){\line(1,0){020}}
    \put(040,260){\line(1,0){020}}
    \put(020,310){\line(1,0){020}}
    \put(000,230){\line(1,0){020}}
    \put(120,250){\line(0,1){020}}
    \put(100,260){\line(0,1){030}}
    \put(100,190){\line(0,1){020}}
    \put(080,280){\line(0,1){030}}
    \put(080,200){\line(0,1){030}}
    \put(060,220){\line(0,1){080}}
    \put(040,260){\line(0,1){100}}
    \put(020,140){\line(0,1){170}}
    \end{picture}
    \end{center}
    \caption{\label{fig:liliales}%
      Two incompatible semi-labeled trees, obtained from study
      S2x4x96c17c14c22 in TreeBASE. The clusters shown with thick lines
      are incompatible.}
    \end{figure}

    A third application of \texttt{Bio::Tree::Compatible} is shown in
    Fig.~\ref{fig:liliales}. The input consists of two semi-labeled trees
    describing the evolution of net-veined Lilliaflorae, obtained from
    study S2x4x96c17c14c22 in TreeBASE.

    Using the \texttt{Bio::Tree::Compatible} module, we have performed a
    systematic study of tree compatibility on TreeBASE, which currently
    contains 2,592 phylogenies with over 36,000 taxa among them. In this
    study, we have found 2,527 pairs of incompatible trees (like those
    shown in Figs.~\ref{fig:angiosperm} to~\ref{fig:liliales}) from a
    total of 3,357,936 pairs of trees. The resulting ratio of 0.075\%
    shows the high internal consistency among the phylogenies, and it
    complements previous large-scale analyses of
    TreeBASE~\cite{piel.ea:2003}.

    \section{Conclusions}
    \label{sect:concl}

    Phylogenetic tree compatibility is the most important concept
    underlying widely-used methods for assessing the agreement of
    different phylogenetic trees with overlapping taxa and combining them
    into common supertrees to reveal the tree of life. The study of the
    compatibility of phylogenetic trees with nested taxa, also known as
    semi-labeled trees, was asked for in
    \cite{page:2004}, and the notion of ancestral
    compatibility was introduced
    in~\cite{daniel.semple:2004,semple.ea:2004}.

    We have analyzed in detail the meaning of the ancestral compatibility
    of semi-labeled trees from the points of view of the local structure
    of the trees, of the existence of embeddings into a common supertree,
    and of the joint properties of their cluster representations. We have
    established the equivalence between ancestral compatibility and the
    absence of certain incompatible pairs and triples of labels in the
    trees under comparison, and have also proved the equivalence between
    ancestral compatibility and a certain property of the cluster
    representations of the trees.

    Our analysis has lead to a very simple polynomial-time algorithm for
    testing ancestral compatibility, which we have implemented and is
    freely available for download from the BioPerl collection of Perl
    modules for computational biology. Future work includes extending the
    \texttt{Bio::Tree::Compatible} implementation into a
    \texttt{Bio::Tree::Supertree} module for building a common supertree
    of two compatible semi-labeled trees.
    \smallskip

\noindent\textbf{Acknowledgements.}
      M. Llabr\'es and F. Rossell\'o have been partially supported by the
      Spanish DGES project BFM2003-00771. G. Valiente was supported by the
      Japan Society for the Promotion of Science through Long-term
      Invitation Fellowship L05511 for visiting JAIST (Japan Advanced
      Institute of Science and Technology). G. Valiente acknowledges with
      thanks R. D. M. Page for many discussions on compatibility of
      phylogenetic trees.

    \bibliographystyle{spmpsci}
    \bibliography{ancestral-jmb-tr}

\begin{thebibliography}{10}
\providecommand{\url}[1]{{#1}}
\providecommand{\urlprefix}{URL }
\expandafter\ifx\csname urlstyle\endcsname\relax
  \providecommand{\doi}[1]{DOI~\discretionary{}{}{}#1}\else
  \providecommand{\doi}{DOI~\discretionary{}{}{}\begingroup
  \urlstyle{rm}\Url}\fi

\bibitem{baum:1992}
Baum, B.R.: Combining trees as a way of combining datasets for phylogenetic
  inference, and the desirability of combining gene trees.
\newblock Taxon \textbf{41}(1), 3--10 (1992)

\bibitem{bininda:2004}
Bininda-Emonds, O.R.P. (ed.): Phylogenetic Supertrees: Combining Information to
  Reveal the Tree of Life, \emph{Computational Biology}, vol.~4.
\newblock Kluwer (2004)

\bibitem{daniel.semple:2004}
Daniel, P., Semple, C.: Supertree algorithms for nested taxa.
\newblock In: O.R.P. Bininda-Emonds (ed.) Phylogenetic Supertrees: Combining
  Information to Reveal the Tree of Life, \emph{Computational Biology}, vol.~4,
  chap.~7, pp. 151--171. Kluwer (2004)

\bibitem{morell:1996}
Morell, V.: Tree{BASE}: The roots of phylogeny.
\newblock Science \textbf{273}(5275), 569--0 (1996).
\newblock \urlprefix\url{http://www.treebase.org}

\bibitem{page:2002}
Page, R.D.M.: Modified mincut supertrees.
\newblock In: Proc. 2nd Int.\ Workshop Algorithms in Bioinformatics,
  \emph{Lecture Notes in Computer Science}, vol. 2452, pp. 537--552.
  Springer-Verlag (2002)

\bibitem{page:2004}
Page, R.D.M.: Taxonomy, supertrees, and the tree of life.
\newblock In: O.R.P. Bininda-Emonds (ed.) Phylogenetic Supertrees: Combining
  information to reveal the tree of life, \emph{Computational Biology}, vol.~4,
  pp. 247--265. Springer-Verlag (2004)

\bibitem{piel.ea:2003}
Piel, W.H., Sanderson, M.J., Donoghue, M.J.: The small-world dynamics of tree
  networks and data mining in phyloinformatics.
\newblock Bioinformatics \textbf{19}(9), 1162--1168 (2003)

\bibitem{ragan:1992}
Ragan, M.A.: Phylogenetic inference based on matrix representation of trees.
\newblock Molecular Phylogenetics and Evolution \textbf{1}(1), 53--58 (1992)

\bibitem{UPC-LSI-04-60-R}
Rossell\'o, F., Valiente, G.: An algebraic view of the relation between largest
  common subtrees and smallest common supertrees.
\newblock Tech. rep., Technical University of Catalonia (2004)

\bibitem{semple.ea:2004}
Semple, C., Daniel, P., Hordijk, W., Page, R.D.M., Steel, M.: Supertree
  algorithms for ancestral divergence dates and nested taxa.
\newblock Bioinformatics \textbf{20}(15), 2355--2360 (2004)

\bibitem{semple.steel:2003}
Semple, C., Steel, M.: Phylogenetics.
\newblock Oxford University Press (2003)

\bibitem{semple.ea:2000}
Semple, C., Steel, M.A.: A supertree method for rooted trees.
\newblock Discrete Applied Mathematics \textbf{105}(1--3), 147--158 (2000)

\bibitem{stajich.ea:2002}
Stajich, J.E., Block, D., Boulez, K., Brenner, S.E., Chervitz, S.A., Dagdigian,
  C., Fuellen, G., Gilbert, J.G., Korf, I., Lapp, H., Lehvaslaiho, H.,
  Matsalla, C., Mungall, C.J., Osborne, B.I., Pocock, M.R., Schattner, P.,
  Senger, M., Stein, L.D., Stupka, E., Wilkinson, M.D., Birney, E.: The
  {B}io{P}erl toolkit: Perl modules for the life sciences.
\newblock Genome Research \textbf{12}(10), 1611--1618 (2002).
\newblock \urlprefix\url{http://www.bioperl.org}

\bibitem{steel.warnow:1993}
Steel, M.A., Warnow, T.: Kaikoura tree theorems: Computing the maximum
  agreement subtree.
\newblock Information Processing Letters \textbf{48}(2), 77--82 (1993)

\bibitem{warnow:1994}
Warnow, T.: Tree compatibility and inferring evolutionary history.
\newblock Journal of Algorithms \textbf{16}(3), 388--407 (1994)

\end{thebibliography}
    \end{document}